\documentclass[apj, numberedappendix]{emulateapj}
\usepackage{verbatim, graphicx, natbib, ifthen}
\newboolean{ColorPlots}
\newcommand{\addgals}{ADDGALS}
\newcommand{\hMsun}{\ h^{-1}\mathrm{M}_{\odot}}

\shorttitle{ArborZ: Photo-$z$s Using Boosted Decision Trees}
\shortauthors{D. Gerdes et al.}

\begin{document}
\setboolean{ColorPlots}{true}

\DeclareGraphicsExtensions{.pdf,.gif,.jpg,.tiff}

\title{ArborZ: Photometric Redshifts Using Boosted Decision Trees}

\author{David W. Gerdes, Adam J. Sypniewski, Timothy A. McKay, Jiangang Hao\altaffilmark{1}, Matthew R. Weis}
\affil{Department of Physics, University of Michigan, Ann Arbor, MI 48109}
\altaffiltext{1}{Present address: Center for Particle Astrophysics, Fermi National Accelerator Laboratory, Batavia, IL 60510}
\email{gerdes@umich.edu}
\author{Risa H. Wechsler, Michael T. Busha}
\affil{Kavli Institute for Particle Physics Astrophysics and Cosmology, Physics Department, and SLAC National Laboratory, Stanford University, Stanford, CA 94305}

\begin{abstract}Precision photometric redshifts will be essential for extracting cosmological parameters from the next generation of wide-area imaging surveys. In this paper we introduce a photometric redshift algorithm, ArborZ,  based on the machine-learning technique of Boosted Decision Trees. We study the algorithm  using galaxies from the Sloan Digital Sky Survey and from mock catalogs intended to simulate both the SDSS and the upcoming Dark Energy Survey. We show that it improves upon the performance of existing algorithms. Moreover, the method naturally leads to the reconstruction of a full probability density function (PDF) for the photometric redshift of each galaxy, not merely a single ``best estimate" and error, and also provides a photo-$z$ quality figure-of-merit for each galaxy that can be used to reject outliers. We show that the stacked PDFs yield a more accurate reconstruction of the redshift distribution $N(z)$. We discuss limitations of the current algorithm and ideas for future work.
\end{abstract}

%\maketitle

\keywords{galaxies: distances and redshifts -- galaxies: statistics -- large-scale structure of the Universe: methods: statistical -- methods: data analysis}

\section{Introduction}
Cosmic expansion makes the redshift of a distant object one of its most fundamental observables. The redshift allows us to estimate distances, and hence to place observed properties (e.g. fluxes) on a physical scale (e.g. luminosities). Whether interpreted as recession velocity or a measure of the change in the scale factor \citep{BunnHogg09}, redshift is defined as the fractional increase in wavelength of the observed spectral energy distribution (SED) $z = \Delta \lambda/\lambda$. As such, it is measured by comparing observed SEDs of distant objects to those of objects nearby, or to atomic and molecular features identified in the lab.

This comparison is relatively straightforward when the two SEDs are both physically similar and well measured, with high signal-to-noise and wavelength resolution adequate to resolve the relevant features. These conditions are often met in spectroscopic surveys, and these typically allow redshifts to be determined with great precision. For example, the $\sim$10$^6$ galaxy redshifts from the Sloan Digital Sky Survey \citep[SDSS;][]{SDSSoverview} have errors of $\Delta z \le 0.0002$. Unfortunately, high resolution spectroscopic data are costly to obtain. Spreading the light from an object into several thousand independent resolution elements typically requires exposures 50-100 times as long as those for broad-band images with the same signal-to-noise. Furthermore, high-resolution spectra imaged on detectors take up much more space than direct images, requiring slit masks, fiber feeds, or image slicers. These challenges have limited the scope of redshift surveys, so that the total number of galaxy spectroscopic redshifts so far measured remains of order a few million.

Many modern astrophysical measurements would benefit from substantially larger catalogs of redshifts,  say 10$^8$ or 10$^9$. These include studies of galaxy evolution, galaxy cluster identification, large scale structure and baryon acoustic oscillation measurements, identification of very high redshift objects, and gravitational lensing studies. Many of these studies would be well served by much more crudely measured redshifts, say $\Delta z \simeq 0.01$. Such redshifts would, for example, allow a 2\% measurement of the distance to a galaxy at $z = 0.5$. Cosmology-induced systematic uncertainty in the conversion from redshift to distance would then dominate uncertainty in determination of the galaxy's properties, making greater accuracy of little benefit for this purpose. For many applications, for example determination of the weak lensing source galaxy distribution, an accurate estimate of the redshift probability density function is as important as the accuracy of the redshift itself \citep{Mandelbaum08}. In many cases, these PDFs are highly non-Gaussian, adding to the complexity of the problem.

It has long been recognized that broadband imaging in several passbands provides a crude measurement of an object's SED \citep{baum62}. In the era of wide-field CCD imaging, precise calibration of broadband photometry is possible, and the low resolution SEDs measured in this way can be used to estimate redshifts. Early efforts to apply photometric redshifts to galaxy evolution \citep{koo81, koo85} and cosmology \citep{lohsp86-2, lohsp86} showed promise, but were limited by the need for precisely calibrated photometry and the lack of adequate spectroscopic training and test catalogs. Use of photometric redshifts has exploded in importance with the onset of massive, well calibrated, multi-band imaging surveys like the SDSS \citep{Oyaizu2008, Lima08, Cunha09}. They have also played an essential role in the study of very deep but smaller-area surveys like GOODS \citep{giav04}, COSMOS \citep{sco07}, and the CFHT Legacy Survey \citep{Coupon09}. Future projects like the Dark Energy Survey \citep{DESoverview} and the Large Synoptic Survey Telescope \citep{ive08} plan to rely heavily on photometric redshifts for central science goals.

One approach to photometric redshift estimation is modeled on the method used for spectroscopic redshift measurement---the comparison of the observed SED to a set of known theoretical or empirical SED templates \citep[e.g. ][]{Fernandez99, Benitez00, Feldmann06}. While this approach can work well, it is complicated by a need for precise understanding of the relative efficiency of the observing system at each wavelength, as well as the need for templates spanning the full range of wavelength and spectral type of the objects observed. An alternative approach is more empirical. These methods begin with a training set of objects for which both photometry in the system of interest and spectroscopic redshifts have been obtained. Ideally, this training set will span the space (in SED, magnitude, and redshift) of the full sample for which photometric redshifts are desired. This training set is then used to define a transformation from points in the multidimensional observed magnitude space to points in a redshift (and possibly SED) space.

Template methods have been favored, and are probably necessary, for estimating redshifts of galaxies inaccessible to spectroscopic redshift determination because they are too faint. As a result, they have played an especially important role in the HDF and UDF \citep{coe06}. For many current and upcoming surveys, the problem is not that spectroscopic redshifts are completely impossible to obtain, but that there are too many objects for which redshifts are required. For these applications, photo-$z$ estimation based on training sets can be very practical. Approaches have also been developed to extend these empirical techniques beyond the limits of available training sets \citep{new08}, and to combine template-based and empirical approaches \citep{Ilbert06, Ilbert09}. Template-fitting methods can easily provide formal fit uncertainties. But since the largest errors occur due to mismatch between the templates and the SED being fit, these errors often significantly underestimate the full uncertainty in photo-$z$ determination. It is likely that full exploration of photo-$z$ uncertainties will require the use of extensive spectroscopic verification sets, which must be kept independent of training sets.

Discovering the mapping between the space of observed magnitudes and redshift-SED space is a classic machine learning problem \citep{mit97}. Many of the approaches familiar in that field have been applied here \citep{Connolly95, Collister2004, Vanzella04, Ball07, Ball08, Carliles08, Freeman09}. Polynomial fitting methods assume a smooth transformation between magnitudes and redshift. These methods have the virtue of simplicity. They are most effective when the parameter space of observables is not too large, the range of SEDs being studied is limited, and the available training sets are extensive. Much more flexibility, and better performance, is possible with methods which allow a more complex mapping. These include a variety of techniques such as local polynomial fitting and artificial neural networks. Another attraction of these methods is that they can easily utilize parameters other than magnitudes, for example galaxy shapes or information about environment, in a natural way. Unfortunately, machine learning methods are particularly unsuited for extrapolating beyond the limits of their training sets; they contain no underlying model to support this.

In this work, we introduce a new machine learning technique to the photometric redshift problem; we estimate photo-$z$s using boosted decision trees (BDTs). A decision tree in its most basic form examines the attributes of a set of data objects to answer a single yes-or-no classification question. A series of sequential cuts is devised to separate the data into one of the two categories. The cuts used on each parameter, and the order in which they are applied, are established using a training set. Performance is tested by running the resulting decision tree on an independent verification set. ``Boosted'' decision trees are developed iteratively. After initial training of the tree, data objects which were originally misclassified are given increased weight, boosting the attention paid to them, and a second tree is generated. This process is iterated tens or hundreds of times, with all the resulting trees combined into a ``forest" to provide significantly enhanced classification power.

For our photo-$z$ determination, we divide the full redshift range into small bins and use a spectroscopic training set to build a set of BDT classifiers for each bin. In essence, each classifier examines each galaxy and evaluates the probability that its redshift falls within the given bin. By examining the distribution of probabilities with redshift we reconstruct a photo-$z$ probability density function for each galaxy. The mean of this distribution provides an excellent ``best estimate" of the photo-$z$, and its shape gives quantitative insight into the actual photo-$z$ PDF.

 This paper begins with a description of the data sets used for training and testing of the BDT algorithm, which we call ArborZ. Section~\ref{sec:BDTdescription} provides a detailed description of the ArborZ approach. Tests of the algorithm on both real SDSS data and simulated data  are then presented in some detail. In both cases, we compare the performance of this method to some other standard methods. 
 \begin{comment}
 We then consider several extensions of the standard ArborZ method, first by adding to the set of input data by including galaxy ellipticity in addition to magnitudes, and then by optimizing the training set for the determination of galaxy cluster redshifts. Both are examples of directions for future work in photo-$z$ estimation. 
 \end{comment}
 We conclude with a review of our results and a summary of ideas for future photo-$z$ projects.

\section{Galaxy Selection}
\label{sec:GalaxySelection}

To train and evaluate the performance of our photometric redshift algorithm, we use real and simulated data from the following sources:

 \textit{SDSS spectroscopic catalog:}  The Sloan Digital Sky Survey \citep[SDSS;][]{SDSSoverview} is an optical imaging survey in the $ugriz$ bands covering $\sim \pi$ steradians of the northern sky. Of the approximately $5\times 10^7$ galaxies detected in this survey, roughly $10^6$ are targeted for spectroscopic \citep{Strauss2002, Blanton2003} followup. We use the catalog from Data
Release 6 \citep{Adelman-McCarthy08}. The spectroscopic sample consists of a magnitude-limited ``main" sample with a median redshift of 0.104 and $r_{Petro}<17.77$, and a subsample of luminous red galaxies (LRGs) that is volume-limited to $z\approx 0.38$ but extends out to $z\approx 0.55$ \citep{Eisenstein01}.  We further enhance this sample by including spectroscopic measurements from other surveys that can be matched to SDSS photometric observations. We include data from the 2dF-SDSS LRG and QSO (2SLAQ) \citep{Cannon06} and the DEEP2 Redshift Survey \citep{Davis03,Davis07}. The final sample consists of 739,000 galaxies. From this sample we reserve 200,000 randomly-selected galaxies for testing, and use the remaining sample for training.
 
 \textit{SDSS mock spectroscopic catalog:} This catalog is derived from a larger mock catalog designed to model the color, magnitude, and spatial distribution of galaxies in the SDSS. The procedure for constructing this catalog is described in Appendix~A. Beginning with this sample, we then apply the spectroscopic selection from the real survey as described above.
  
   \textit{Dark Energy Survey mock catalog:} The Dark Energy Survey \citep[DES;][]{DESoverview} is a planned 5000-square-degree survey of the southern sky using the Blanco 4-meter telescope at CTIO. The five-year survey will collect image data in five optical passbands, $grizY$, over 520 nights beginning in 2011. Photometric redshifts for this survey have previously been studied by \citealt{Lin04} and \citealt{Banerji08}. The 573-square-degree mock catalog consists of $\sim 2\times 10^7$ galaxies with $z<1.4$, with photometry intended to replicate the five-year sensitivity of the DES. The sample, also constructed as described in Appendix~A, is magnitude-limited to $r< 24$. We use a sample of $\sim$ 510k randomly-selected galaxies for training, and another randomly-selected 200k galaxies for testing, which provide adequate statistics for this study.

\begin{figure}[htbp]
\begin{center}
\includegraphics[height=0.45\columnwidth, angle=90]{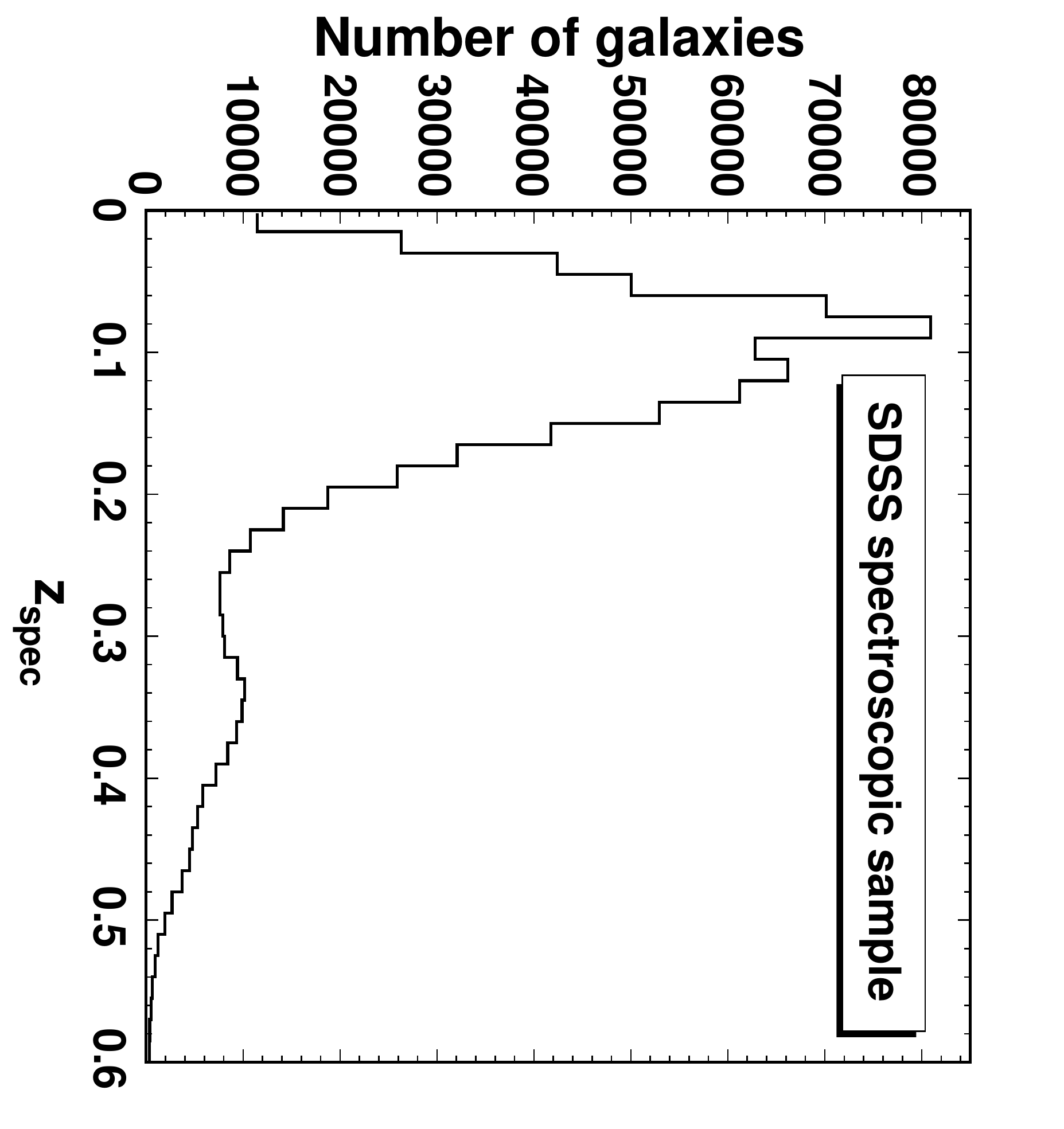}
\includegraphics[height=0.45\columnwidth, angle=90]{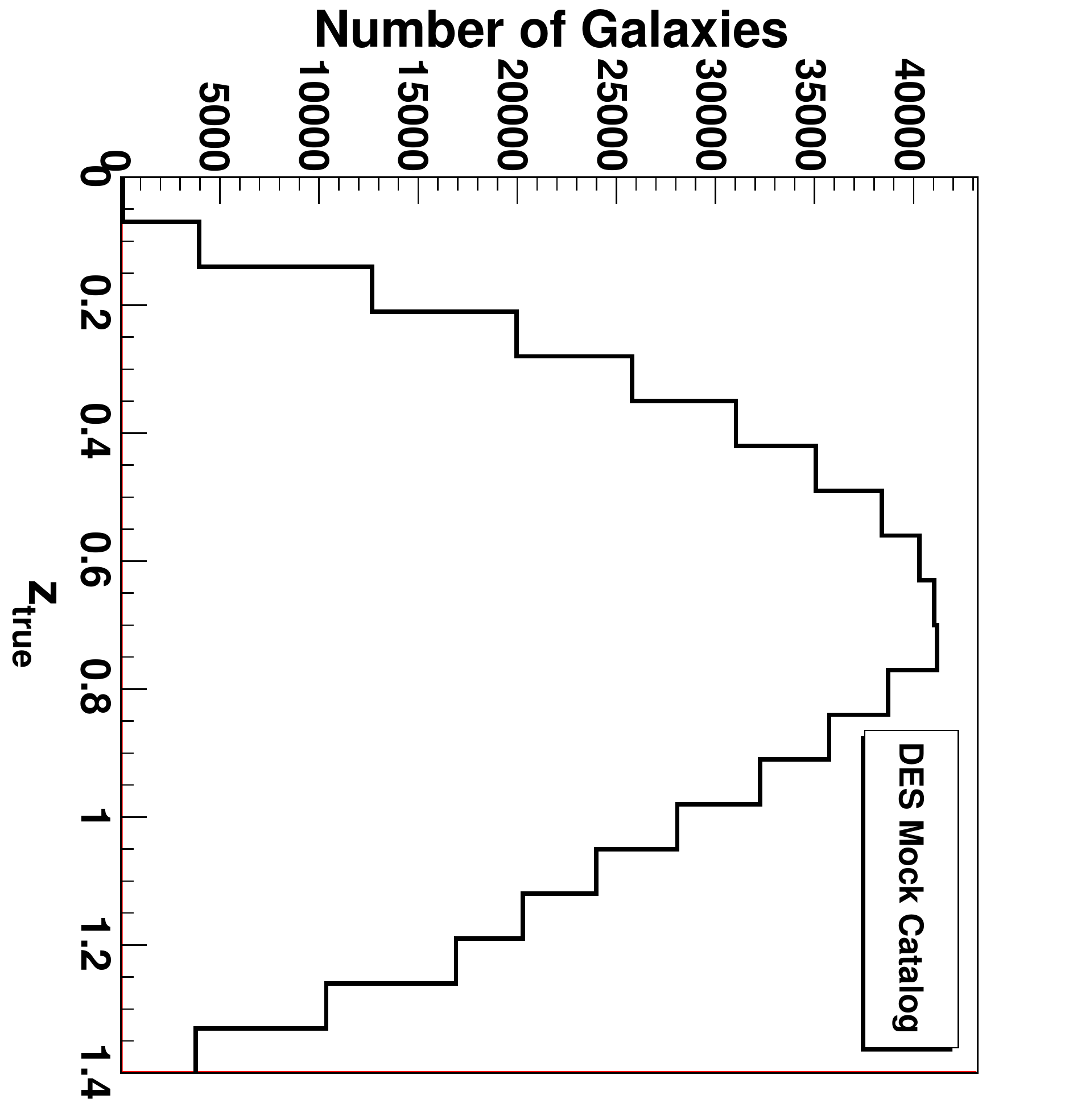}
\caption{\label{fig:training-zdists}Redshift distribution for the SDSS spectroscopic sample (left) and the DES mock catalog (right). The SDSS sample shows two peaks, one near $z=0.1$ for the magnitude-limited sample, and a smaller one near $z=0.35$ for the volume-limited sample of luminous red galaxies. The DES mock catalog is magnitude-limited to $r<24$.}
\end{center}
\end{figure}

The redshift distributions for these samples are shown in Figure~\ref{fig:training-zdists}.

\section{The Boosted Decision Tree Photo-$z$ Algorithm: ArborZ}
\label{sec:BDTdescription}

Consider the general problem of classifying a set of objects, characterized by a vector of observable variables $\mathbf{x}$, into two different populations: ``signal" or ``background". When the two populations are relatively disjoint, simple cuts may be sufficient to achieve good efficiency with high purity. More realistic and complex situations require more sophisticated approaches, such as machine-learning techniques. Boosted decision trees (BDTs) \citep{Hastie2001} are one of the most successful such techniques to emerge in recent years, and have found applications in areas as diverse as text recognition \citep{Howe2005}, spam filtering \citep{Drucker1999}, and particle identification in high-energy physics \citep{Roe2005}. 

To adapt a binary classifier to the problem of assigning a continuous photometric redshift, we divide the spectroscopic training set into a series of redshift bins, $\Delta z_i$. Each bin is assigned its own BDT classifier. The $N$ galaxies whose redshifts fall into bin $i$ form the ``signal" training set for the $i$th classifier.  To form the corresponding background training set, we choose $5N$ galaxies at random from the set of all galaxies whose spectroscopic redshifts fall more than 3$\sigma$ away from the bin in question, where $\sigma$ is the approximate expected resolution of the photo-$z$ algorithm in the target sample ($\sigma = 0.02$ for SDSS data). This $3\sigma$ cut provides a clean separation between the signal and background training sets, preventing the algorithm from overtraining itself by trying to distinguish objects that are nearly identical to within errors. The choice of $5N$ galaxies for the background training set helps enhance the training statistics. Each galaxy in the target evaluation set is then examined by the ensemble of classifiers, and the resulting distribution of probabilities can be used to extract either a single best-estimate photo-$z$ or converted into a photo-$z$ probability density function. We use the boosted decision tree algorithms implemented in the Toolkit for Multivariate Analysis \citep{Hocker2007}.

\subsection{Description}

The process of training a boosted decision tree classifier begins with the construction of a single decision tree. First, a root node is formed containing all the objects. Each object has a weight $w_i$ that is initially set to unity. The root node is then split into a left branch and a right branch by placing a cut on the one variable that gives the best separation between signal and background.
To determine the optimal cut to split a node, we define the purity in a given branch by
$$
	p = \frac{\sum_s w_s}{\sum_s w_s + \sum_b w_b},
$$
where $w_s$ and $w_b$ are the weights of the signal and background objects, respectively. We then define the Gini index \citep{Breiman1984},
$$
	G = \left(\sum_{i=1}^n w_i\right) p(1-p),
$$
where $n$ is the number of objects on that branch. The split is made by scanning over the range of each variable and determining which cut on which variable maximizes the increase in the Gini index between the parent node and the sum of the Gini indices of the left and right branches. This splitting process is repeated until some stopping criterion is reached, for example a minimum number of objects on each leaf. A terminal leaf with a purity above some given threshold is called signal leaf; otherwise it is a background leaf. By construction each object falls on either a signal or a background leaf.

Individual decision trees are relatively weak classifiers. Furthermore, small fluctuations in variables with similar discriminating power can lead to quite different tree structures, with possibly different discriminating ability. The boosting procedure allows an ensemble of such weak classifiers to be combined into a single, powerful classifier. We use the AdaBoost algorithm of Freund and Schapire \citep{Freund1997}. In this procedure, we iteratively generate new decision trees by assigning a higher weight to objects that were previously misclassified. The misclassification rate $R$ of a given tree is defined by
$$
	R = 1 - \max(p, 1-p).
$$
The subsequent tree is trained by ``boosting" the weight of each misclassified object by a factor
$$
	\alpha = \frac{1-R}{R},
$$
and then rescaling the weights of all the objects to keep $\sum w_i$ the same for all the trees.
This process is repeated many times, resulting in a ``forest" of trees. We find that forests of 50-100 trees give good results, with little improvements from larger forests.

As a final step to prevent overtraining, the trees are pruned to remove statistically insignificant nodes. Define the cost complexity \citep{Breiman1984} $\rho$ for a given node to be
$$
     \rho = \frac{R(\mathrm{node})-R(\mathrm{subtree~below~that~node})}{\#\mathrm{nodes(subtree~below~that~node)}  -1}.
$$
We iteratively remove the node with the smallest $\rho$ value as long as $\rho$ is less than some pruning-strength threshold; we obtain the best results with $\rho=4.5$.  Any duplicate trees that remain after the pruning step are removed.

 The final score of an object is a weighted sum of its score on each tree in the forest:
 $$
           y(\mathbf{x}) = \sum_{\mathrm{trees}~i} \ln\alpha_i\cdot h_i(\mathbf{x}),
 $$
 where the subscore $h_i$ on an individual tree is $+1$ if the object is classified as signal and $-1$ if it is classified as background. This procedure gives higher weight to trees with lower misclassification rates. The more signal-like an object appears, the larger its score.

\begin{figure}[htb]
\begin{center}
\includegraphics[height=0.7\columnwidth, angle=90]{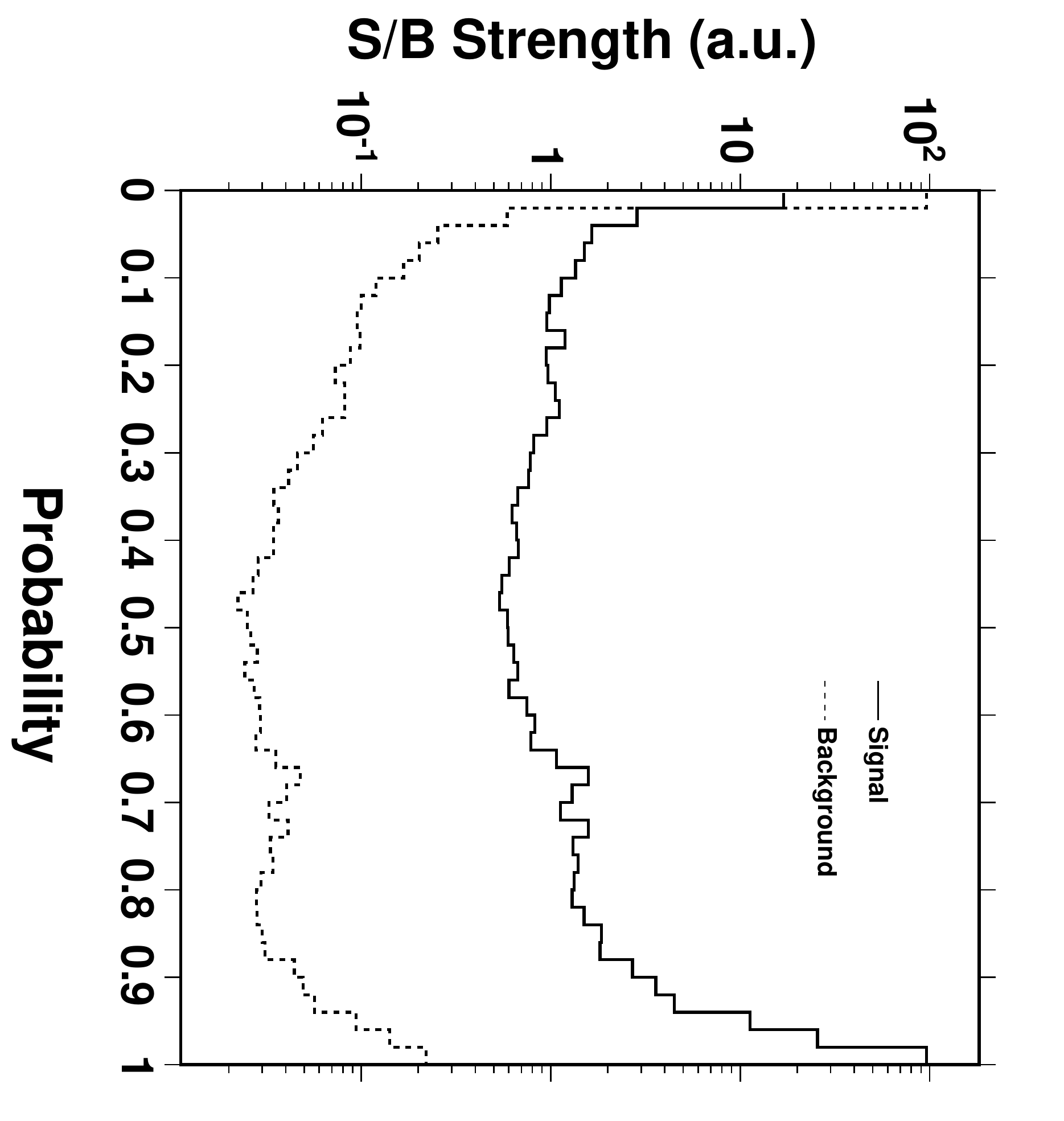}
\includegraphics[height=0.7\columnwidth, angle=90]{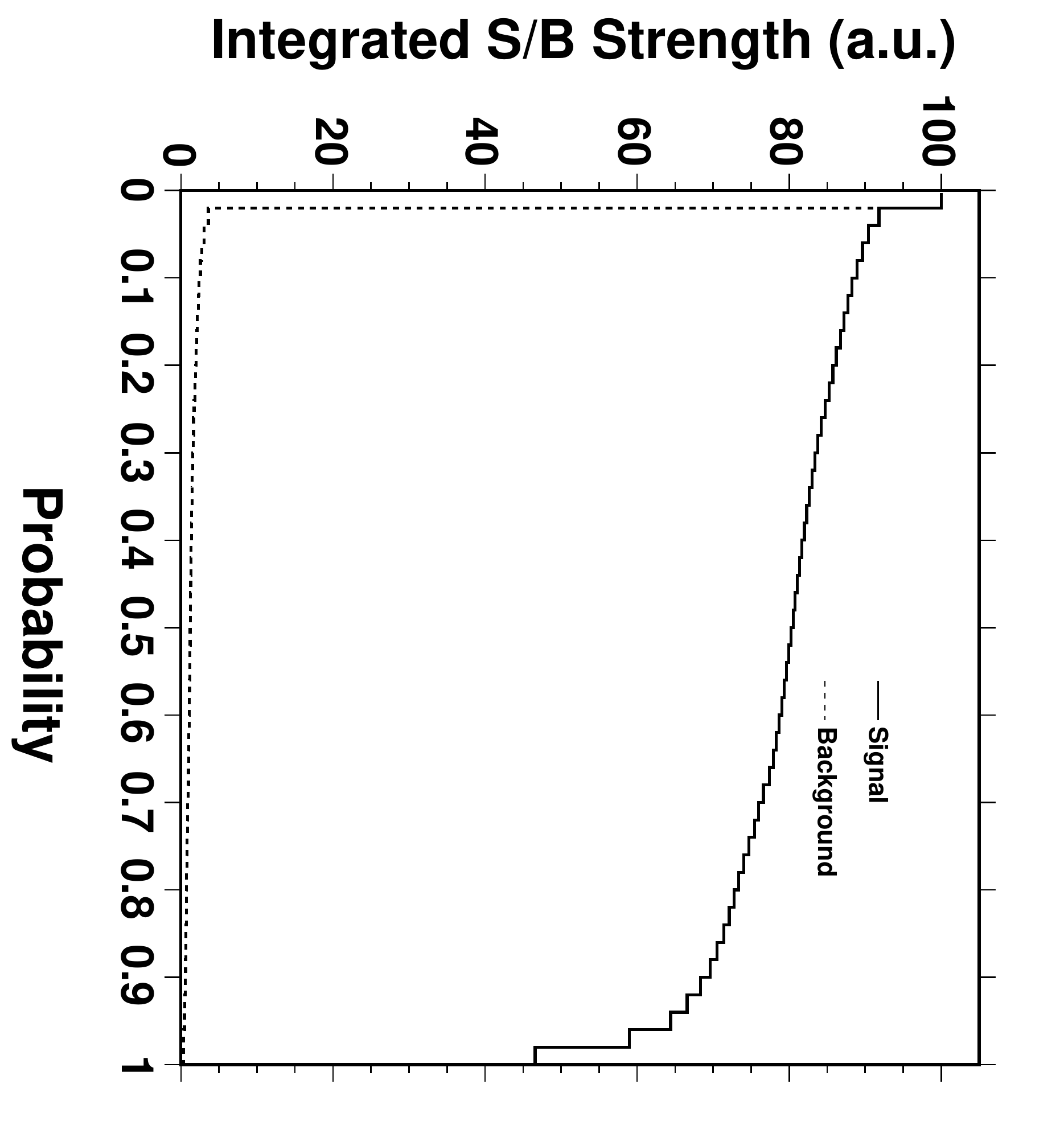}
\caption{\label{fig:SBproba} Top: Classification probability distributions  for signal and background galaxies, normalized to equal areas. Bottom: Integrated classification probability distributions, showing the fraction of signal and background galaxies with probability greater than the given value. }
\end{center}
\end{figure}

 The distribution of scores for signal and background objects can be converted into signal and background classification probabilities, $\hat{y}_{S(B)}$.  These distributions are shown in Figure~\ref{fig:SBproba} for galaxies in the SDSS spectroscopic evaluation sample. Background galaxies---that is, those whose spectroscopic redshifts fall more than $3\sigma$ outside the signal region in question---have probabilities strongly peaked at low values. Thus, the BDT classifier probability is a strong redshift discriminator.
Finally, knowing these classification probabilities, we can compute the probability that a galaxy with a given BDT score falls in the signal redshift bin:
 $$
   P_{S,i} = \frac{f_S\cdot \hat{y}_{S,i}}{f_S\cdot \hat{y}_{S,i}+(1-f_S)\cdot\hat{y}_{B,i} },
 $$
 where the signal fraction $f_S = N_S/(N_S + N_B)$ is the expected fraction of galaxies in each redshift bin, and $N_S$, $N_B$ are the expected numbers of signal and background galaxies. The expected signal fraction in each redshift bin can be obtained directly from the redshift distribution of the training set, in cases where the redshift distribution in the target evaluation set is similar. In fact, as seen in Figure~\ref{fig:SBproba}, the signal and background classification probabilities are sufficiently well-separated that the signal probability $P_S$ is relatively insensitive to the choice of signal fraction. On the other hand, any training-set-based method will have difficulty when the target evaluation set's properties---whether magnitudes, colors, or redshifts---differ substantially from those of the training set.

\subsection{Performance in SDSS Data}
\label{sec:performance-SDSS}

 The SDSS spectroscopic training sample of 510k galaxies described in Section~\ref{sec:GalaxySelection} is divided into 64 redshift bins containing equal numbers of galaxies. This ensures that each training subsample has equal statistics; as a result, the redshift bins themselves vary in width. The $N$ galaxies in bin $i$ form the signal training set for the $i$th classifier, while the background training set consists of $5N$ galaxies chosen at random from the set of galaxies with redshifts at least $\Delta z = 0.06$ (approximately $3\sigma_{z,photo}$) away from the redshift bin in question.  We use the five observed magnitudes $ugriz$ as our training variables.
  
\begin{figure}[htbp]
\begin{center}
\includegraphics[height=0.45\columnwidth, angle=90]{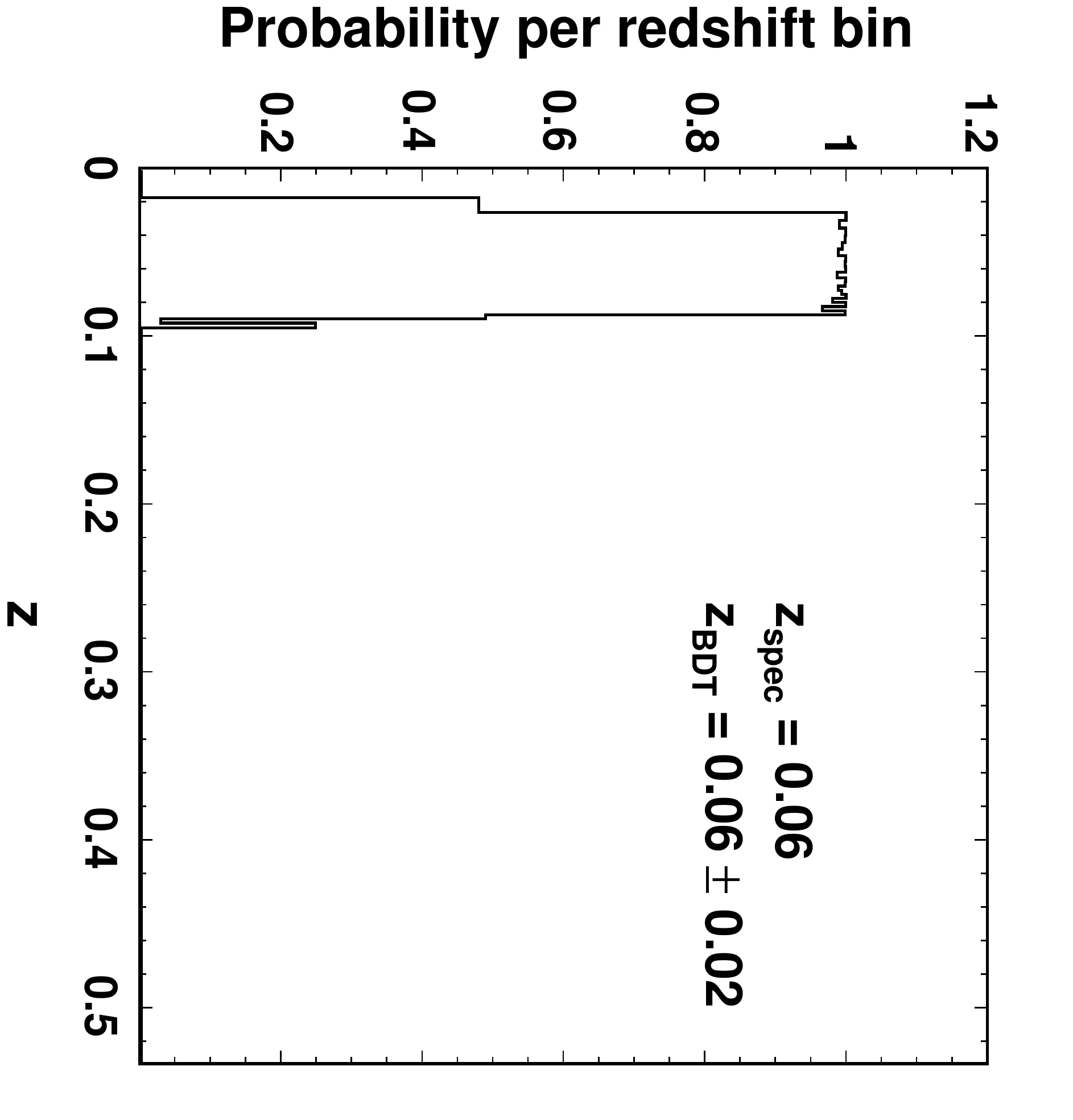}
\includegraphics[height=0.45\columnwidth, angle=90]{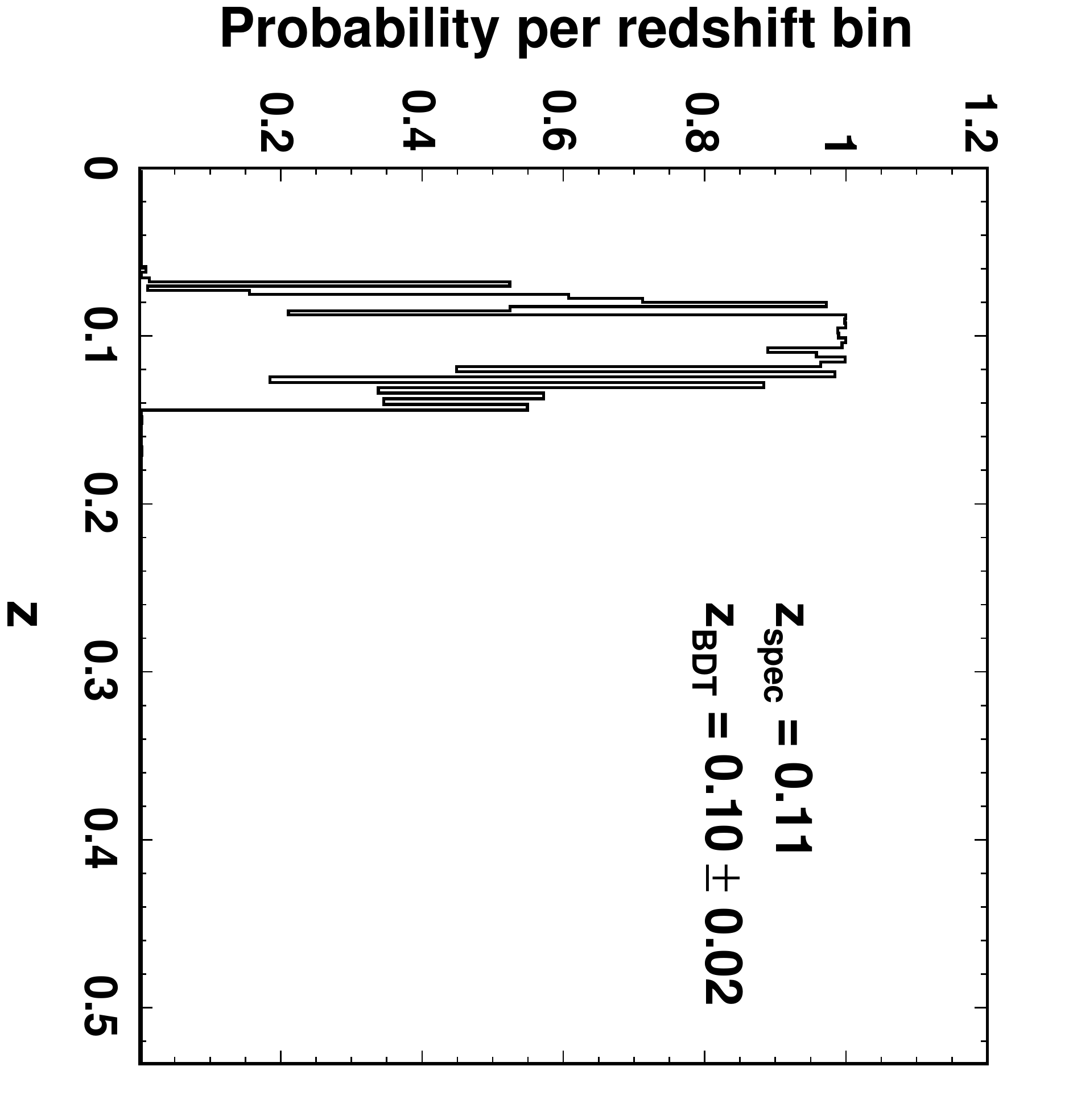}
\includegraphics[height=0.45\columnwidth, angle=90]{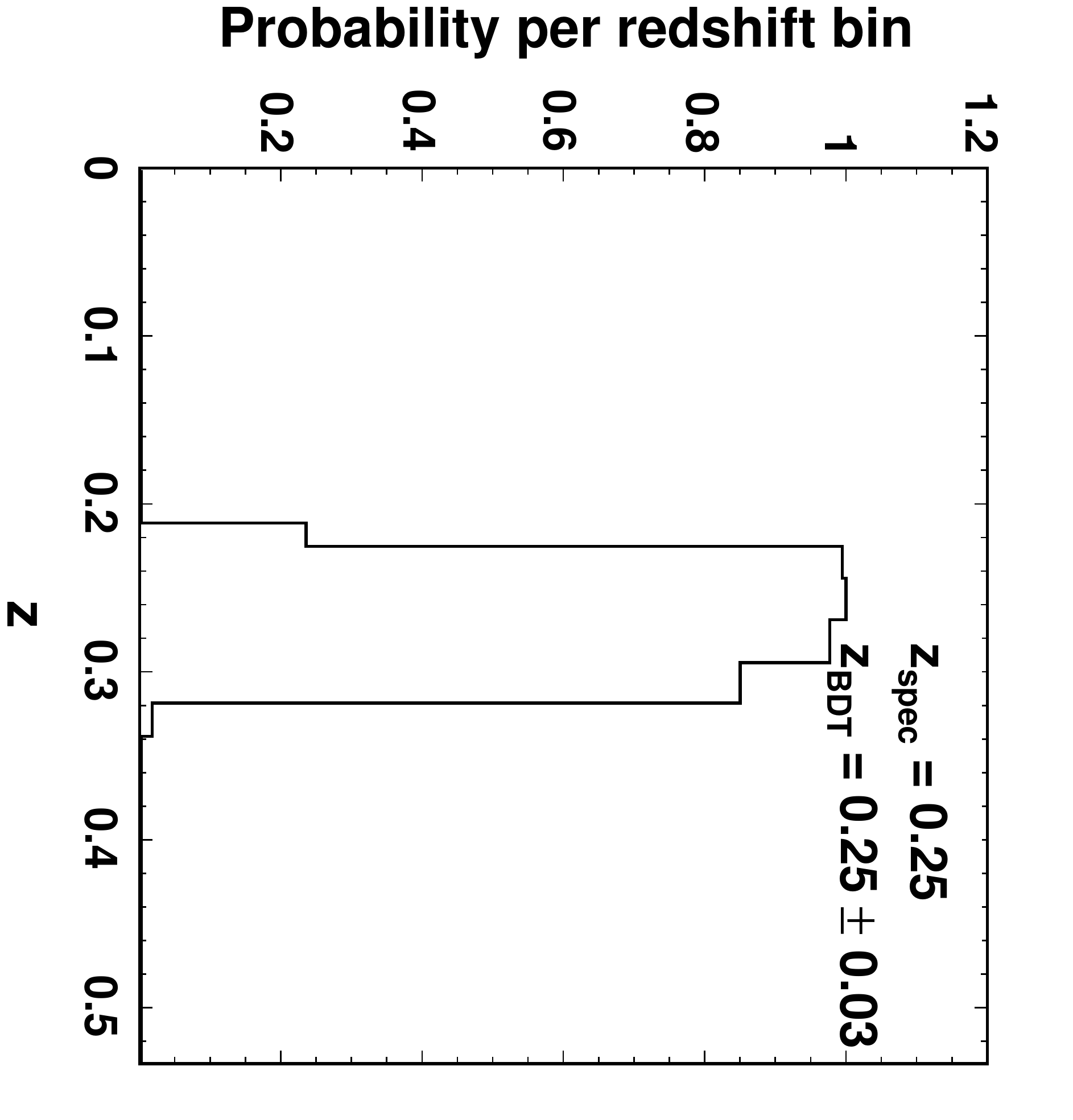}
\includegraphics[height=0.45\columnwidth, angle=90]{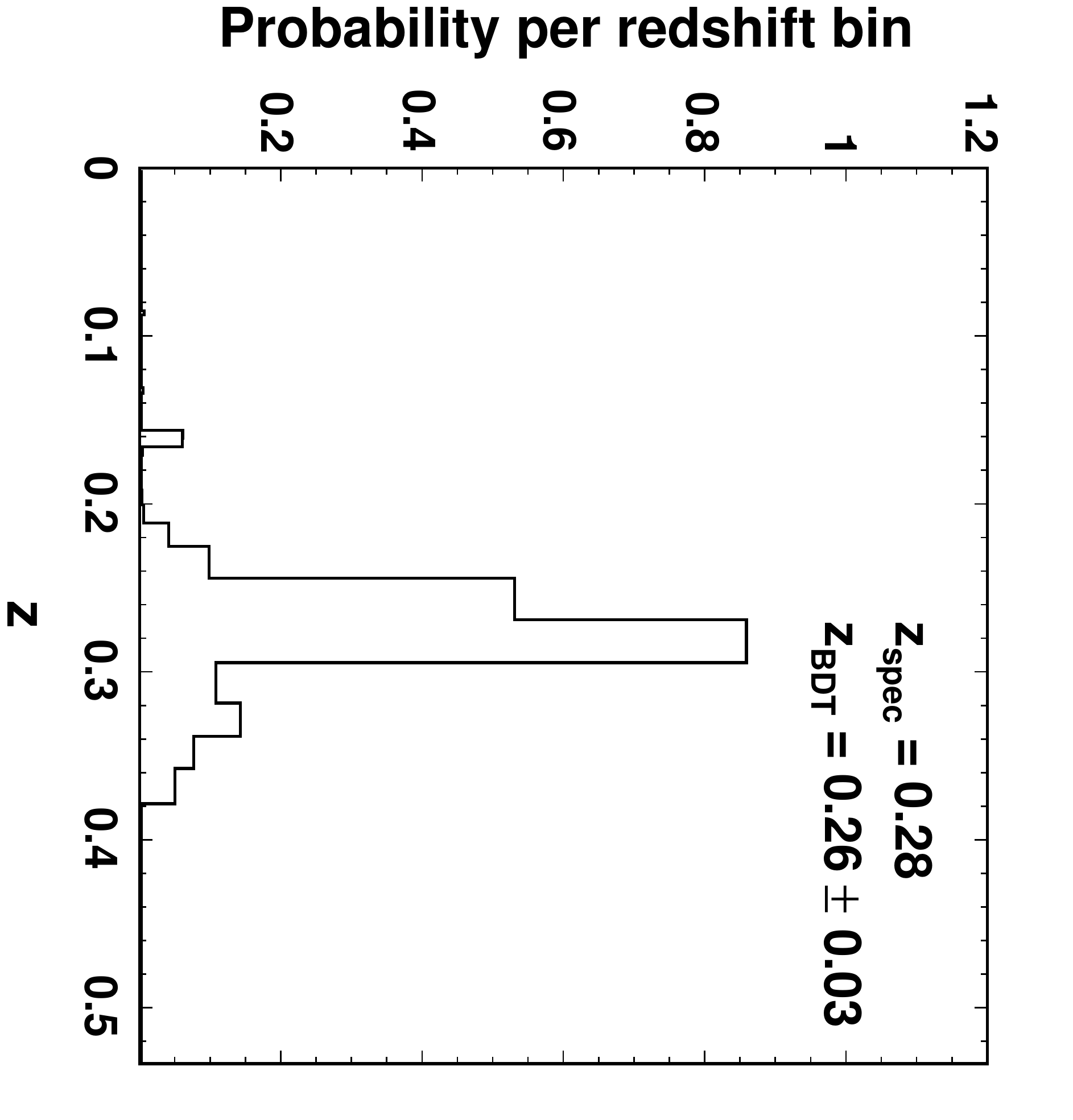}
\includegraphics[height=0.45\columnwidth, angle=90]{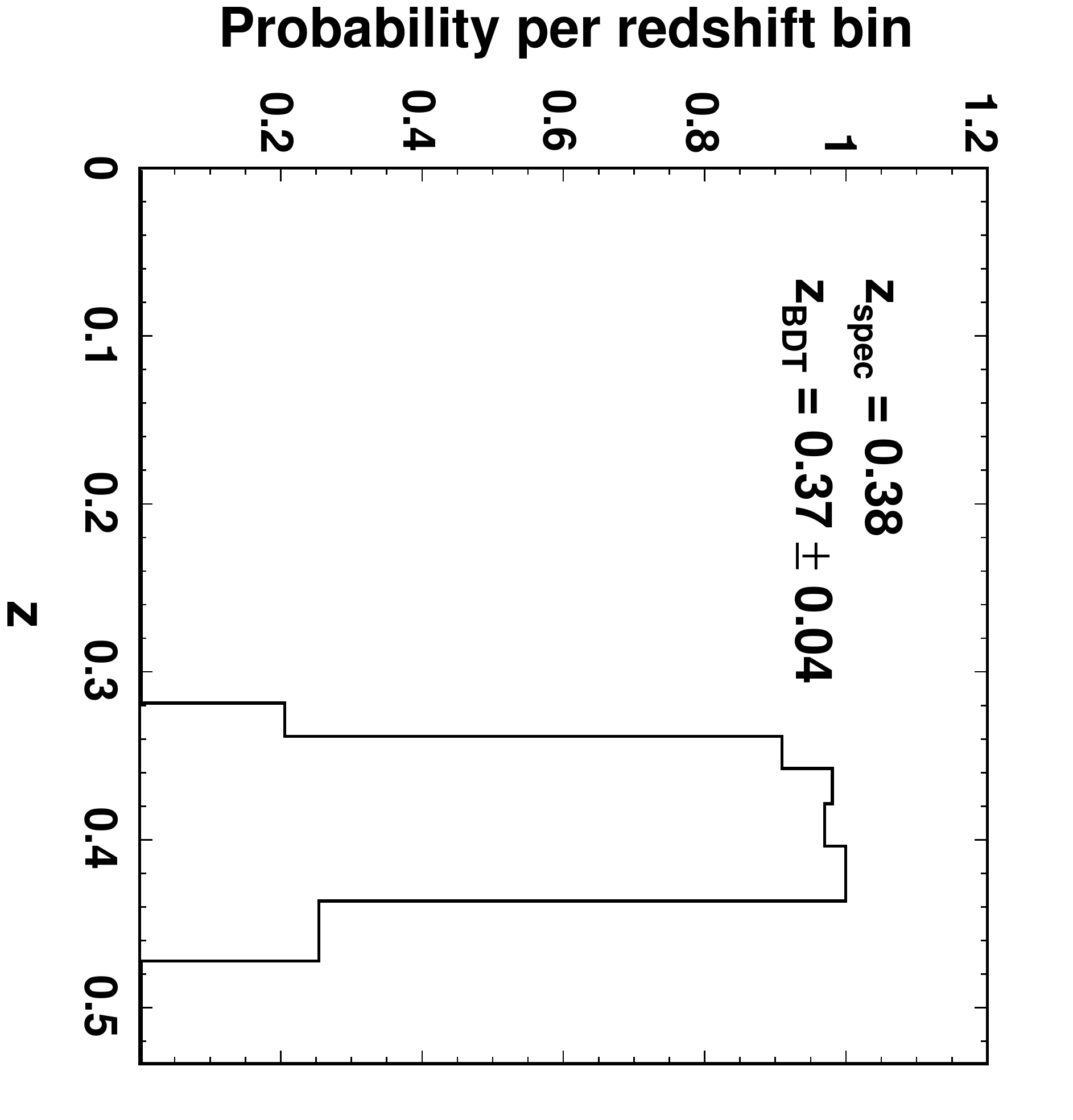}
\includegraphics[height=0.45\columnwidth, angle=90]{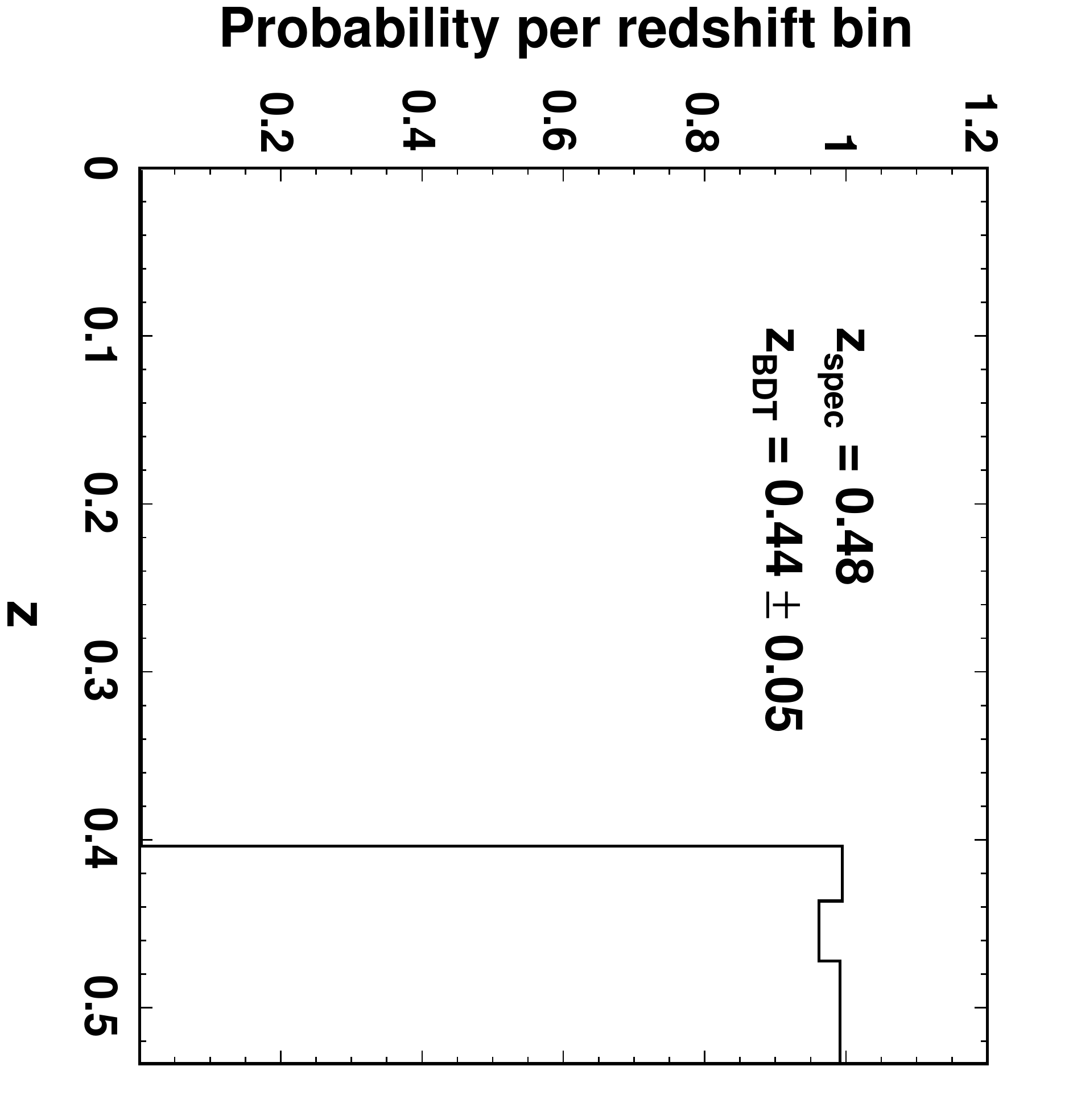}

\caption{\label{fig:BDToutputs} Distribution of ArborZ boosted decision tree classifier probabilities as a function of redshift for some individual galaxies in the SDSS spectroscopic evaluation set. }
\end{center}
\end{figure}

 The algorithm's performance is evaluated on a subsample of 200,000 galaxies that were excluded from the training process. The ensembles of classifier probabilities for each redshift bin, $P_{S,i}$ for some typical galaxies are shown in Figure~\ref{fig:BDToutputs}. The mean of this histogram determines the ``best-estimate" photo-$z$ for each galaxy, and the range containing the middle 68\% of the area determines the error. However, the full power of the algorithm comes from the reconstruction of the complete probability distribution itself.  

\begin{figure}[htbp]
\begin{center}
\ifthenelse{\boolean{ColorPlots}}
{
  \includegraphics[height=\columnwidth, angle=90]{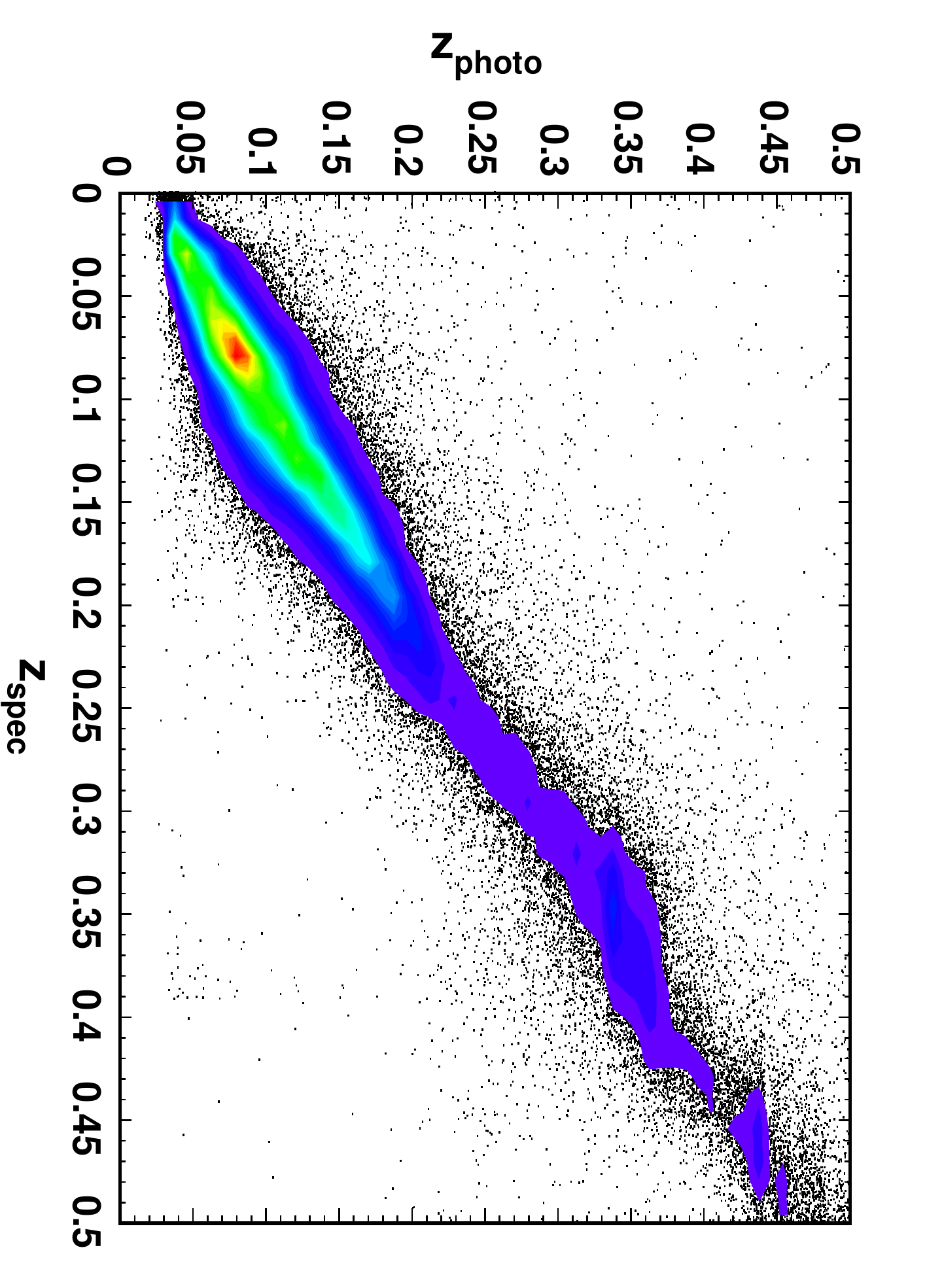}
}
{
\includegraphics[height=\columnwidth, angle=90]{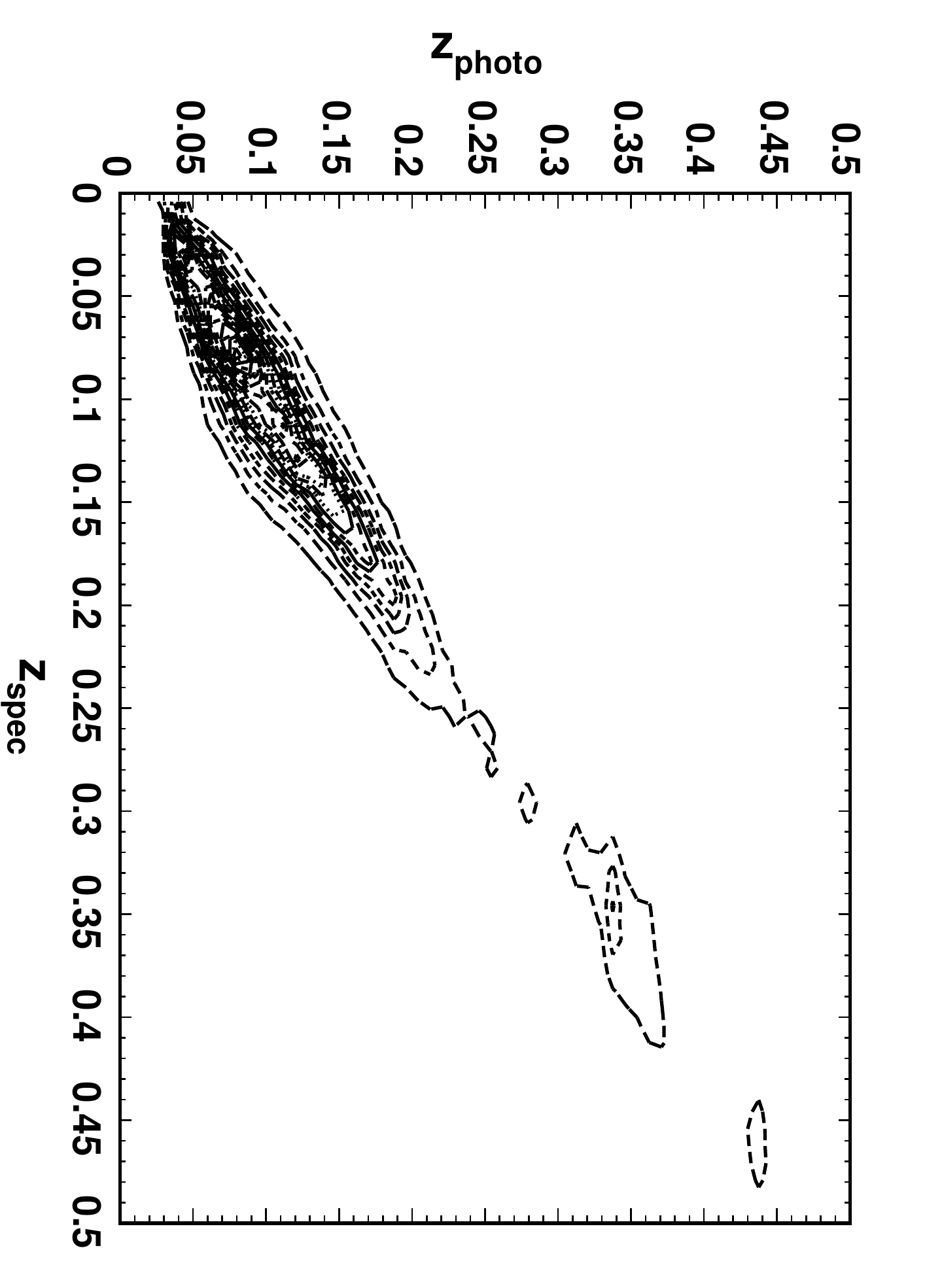}
}
\caption{\label{fig:BDTscatt} ArborZ photo-$z$ estimate vs. spectroscopic redshift for the SDSS spectroscopic evaluation set.}
\end{center}
\end{figure}

\begin{figure}[htbp]
\begin{center}
\ifthenelse{\boolean{ColorPlots}}
{\includegraphics[height=\columnwidth, angle=90]{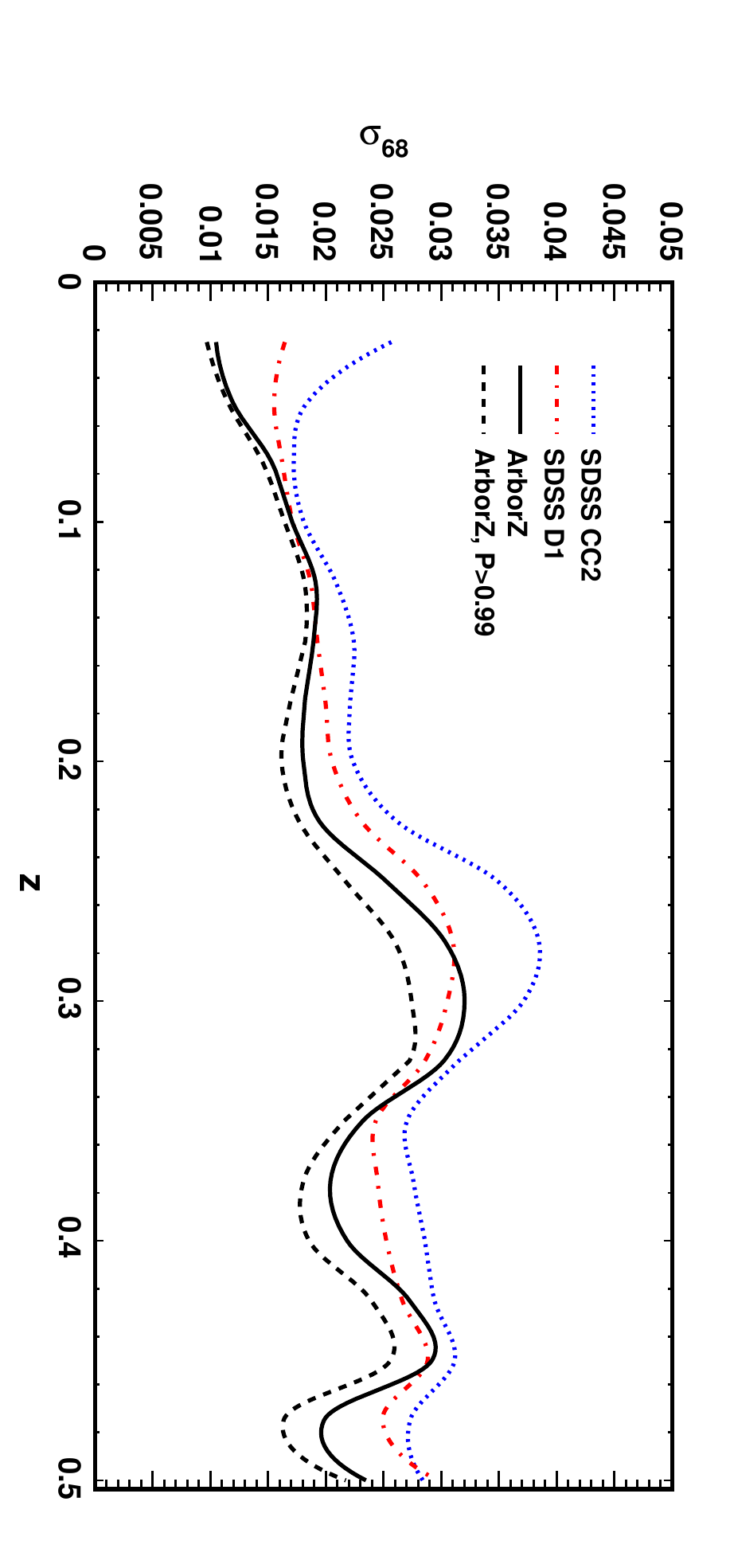} }
{\includegraphics[height=\columnwidth, angle=90]{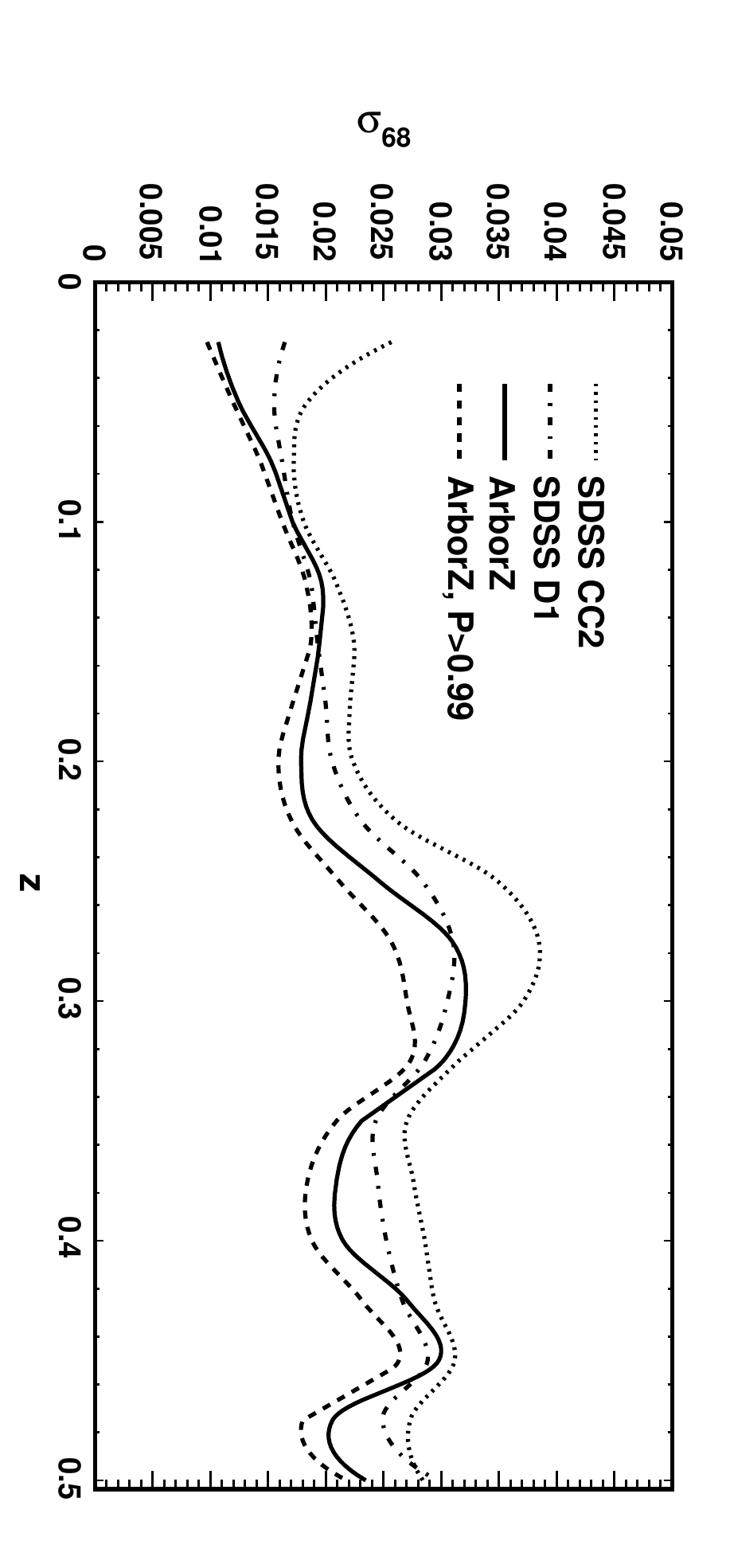} }
\caption{\label{fig:BDTpulls} ArborZ performance in the SDSS spectroscopic evaluation set, compared with the SDSS production photo-$z$ algorithms D1 and CC2. The plot shows the 68\% confidence-interval width of the photo-$z$ residual distribution $z_{photo}-z_{spec}$ as a function of spectroscopic redshift. The dashed ArborZ curve shows the effect of placing a cut on the peak classifier probability as described in \protect\ref{sec:Errors}. This cut rejects 11\% of the galaxies.}
\end{center}
\end{figure}

The photometric redshift obtained from the distribution of BDT probabilities for the evaluation set is shown as a function of spectroscopic redshift in Figure~\ref{fig:BDTscatt}. Figure~\ref{fig:BDTpulls} shows the 68\% confidence-interval width of the  residual distribution $z_{photo}-z_{spec}$ as a function of spectroscopic redshift. For comparison, we also show the performance of the two production SDSS photo-$z$ algorithms, D1 and CC2 \citep{Oyaizu2008}. These methods both employ neural networks, where the training variables are the $ugriz$ magnitudes and $u-g$, $g-r$, $r-i$, and $i-z$ colors respectively. The ArborZ algorithm equals or exceeds the performance of these two algorithms over most of the redshift range in the sample.

\subsection{Error Estimation}
\label{sec:Errors}

\begin{figure}[htbp]
\begin{center}
\includegraphics[height=0.7\columnwidth, angle=90]{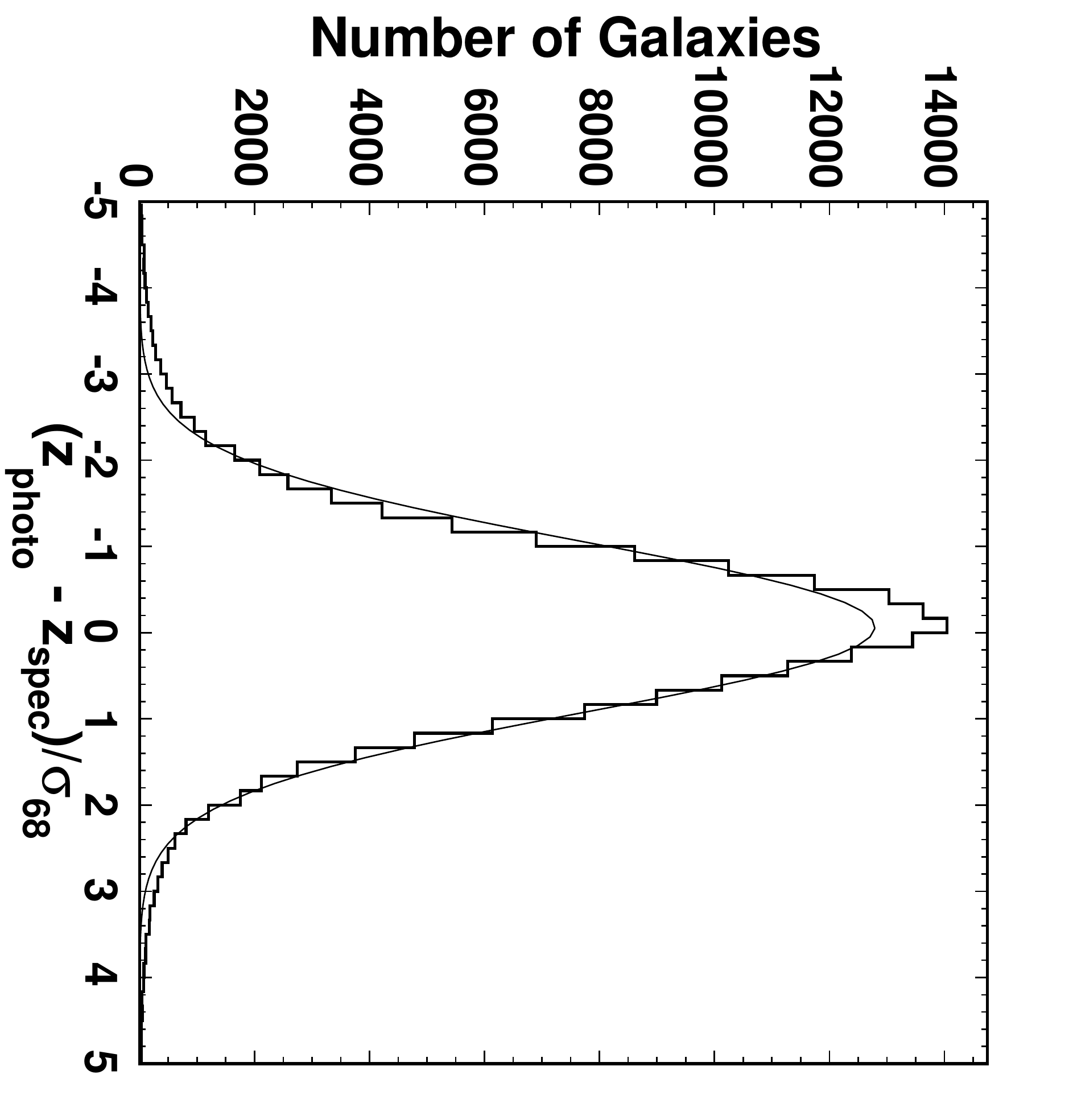}
\caption{\label{fig:pullnorm} Normalized residual distribution $(z_{photo}-z_{spec})/\sigma_{68}$ for galaxies in the SDSS spectroscopic evaluation sample. The fit is to a Gaussian with a mean of $-0.006$ and a width of $0.985$. }
\end{center}
\end{figure}

The ArborZ galaxy-by-galaxy probability distributions such as those shown in Figure~\ref{fig:BDToutputs} provide several different methods to estimate the photo-$z$ error. First, as noted above, we can simply determine the width of the region of the probability distribution that contains the middle 68\% of the area, $\sigma_{68}$.  This is our default method. Figure~\ref{fig:pullnorm} shows the normalized residual distribution, $(z_{photo}-z_{spec})/\sigma_{68}$, in the SDSS evaluation sample. This distribution is well-described by a gaussian with a mean of $-0.006$ and a width of $0.985$, indicating that the errors are properly estimated and that the photo-$z$s are unbiased.  The fraction of catastrophic mismeasurements, which we define to be cases where $z_{photo}-z_{spec}>3\sigma_{z_{photo}}$, is 1.9\%, compared to 2.6\% (3.7\%) for the D1 (CC2) photo-$z$ algorithm. However, the D1 and CC2 algorithms have normalized residual widths of 1.12 and 1.13 respectively, so these algorithms may be underestimating their errors by roughly 10\%. 

\begin{figure}[htb]
\begin{center}
\includegraphics[width=\columnwidth]{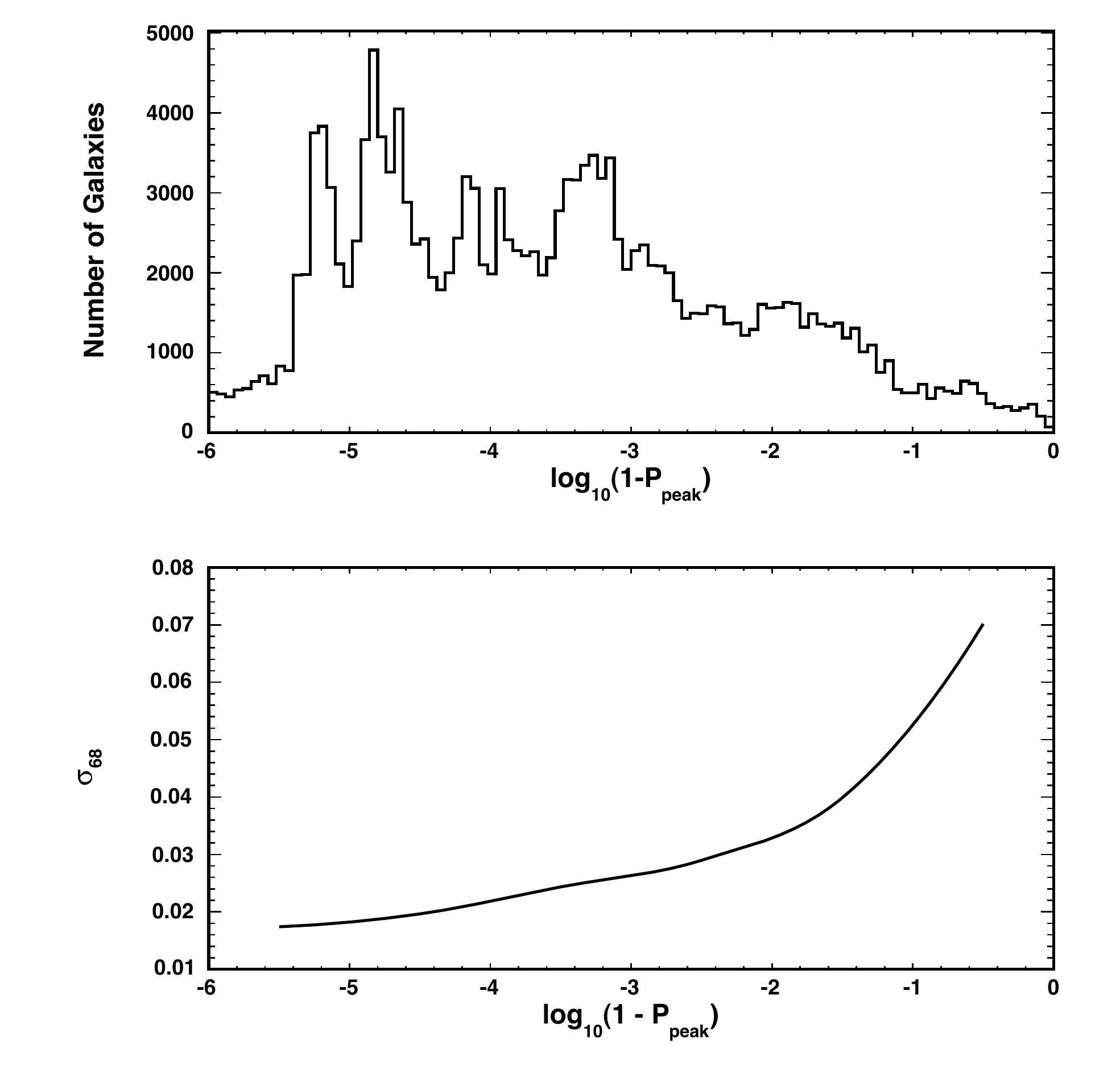}
\caption{\label{fig:peakBDTscore} Top: Peak BDT probability distribution in the SDSS spectroscopic evaluation set. Bottom: Photo-$z$ error as a function of peak probability.}
\end{center}
\end{figure}

An alternate approach to estimating the photo-$z$ error exploits the fact that the BDT probabilities provide a quantitative figure-of-merit for the classification strength: the more signal-like an object appears to a given classifier, the higher its probability for that redshift bin. We therefore expect galaxies with larger peak probabilities ($P_{peak}$)  to be more reliably measured. The distribution of peak probabilities in the SDSS spectroscopic evaluation sample is shown  in the top panel of Figure~\ref{fig:peakBDTscore}. 
In the bottom panel, we show the width of the residual distribution $z_{photo}-z_{spec}$ as a function of the peak probability, which displays the expected correlation.

In applications where well-measured photo-$z$s are a prime concern and some reduction in statistics can be tolerated, one could place a cut on the peak probability to obtain a better-measured subsample of galaxies. 
For example, requiring that the peak probability be greater than 0.99 (0.90) retains 88\%  (99\%) of the galaxies in the SDSS spectroscopic sample. Galaxies that pass the $P_{peak}>0.99$ cut have a mean photo-$z$ error of 0.021. The mean photo-$z$ error of galaxies rejected by this cut is 0.041, nearly twice as large.

\subsection{Reconstructed Redshift Distribution $N(z)$ and the Photo-$z$ PDF}
\label{sec:PDF}

The methods described above provide a single best estimate of each galaxy's photometric redshift, together with an estimated Gaussian error, as is the common practice for most photo-$z$ algorithms. Such estimates, however, are generally biased \citep{Lima08}. The BDT apparatus, with its evaluation of the classification probability for each redshift, leads naturally to the reconstruction of each galaxy's full photo-$z$ probability density function. For many applications, such as measurements of weak gravitational lensing or galaxy-galaxy correlations for baryon acoustic oscillation surveys, individual galaxy-by-galaxy photo-$z$s are less important than an accurate count of the number of galaxies in each redshift bin, $N(z)$. For this purpose, the PDFs are more useful and less biased than the best-estimate photo-$z$s.

\begin{figure}[htb]
\begin{center}

\ifthenelse{\boolean{ColorPlots}}
{ 
\includegraphics[width=\columnwidth]{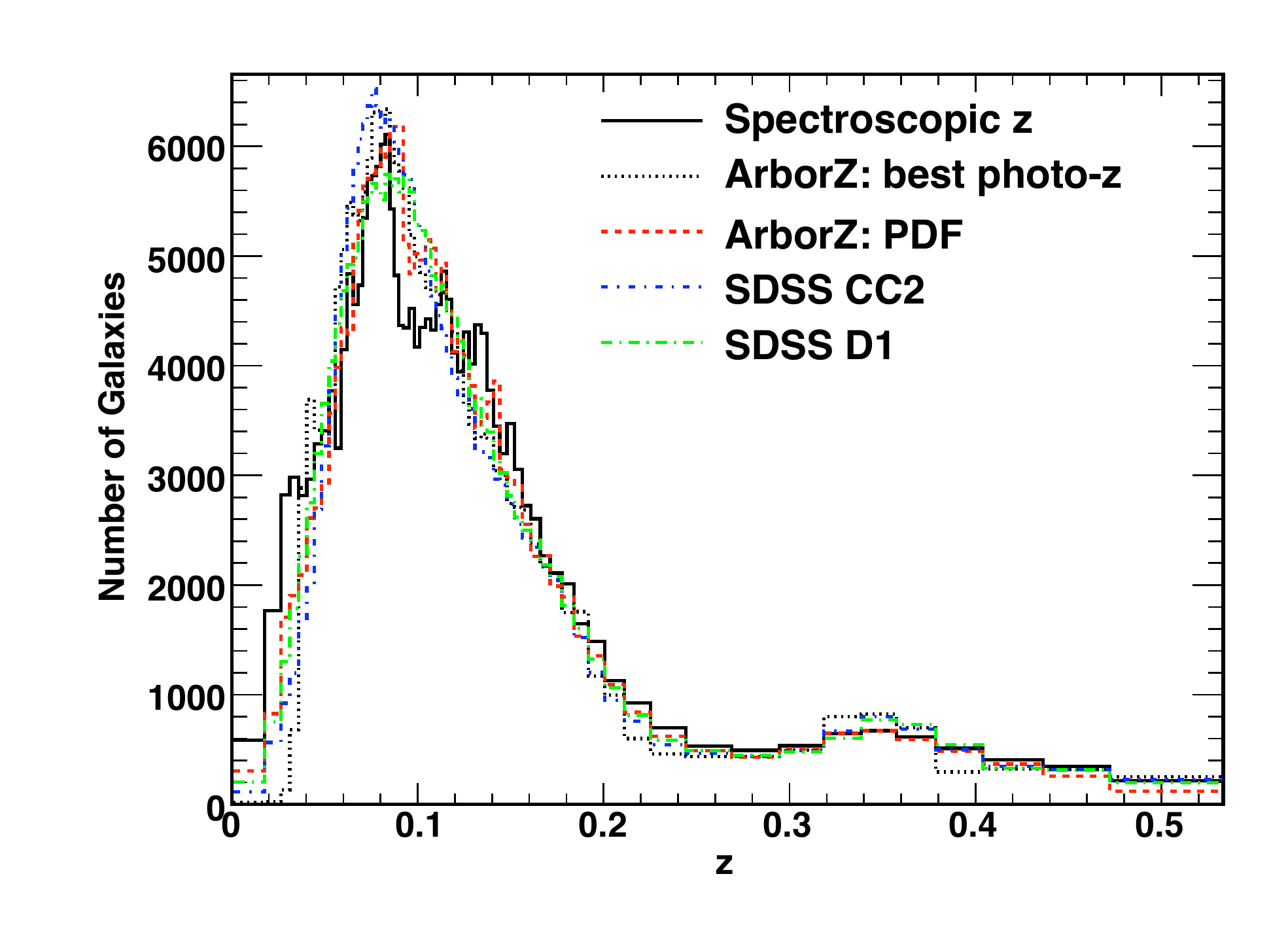} 
\includegraphics[height=\columnwidth,angle=90]{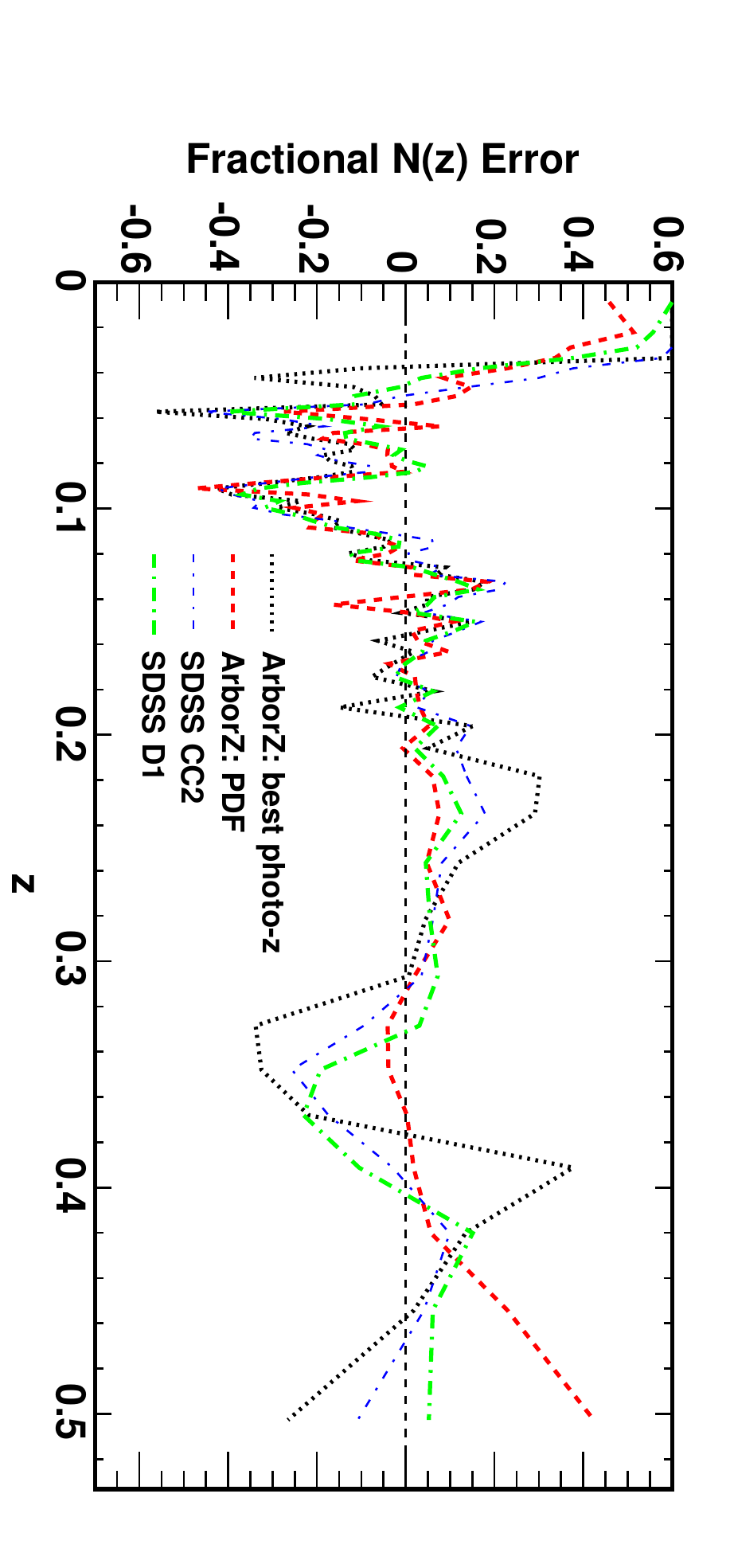} 
}
{
\includegraphics[width=\columnwidth]{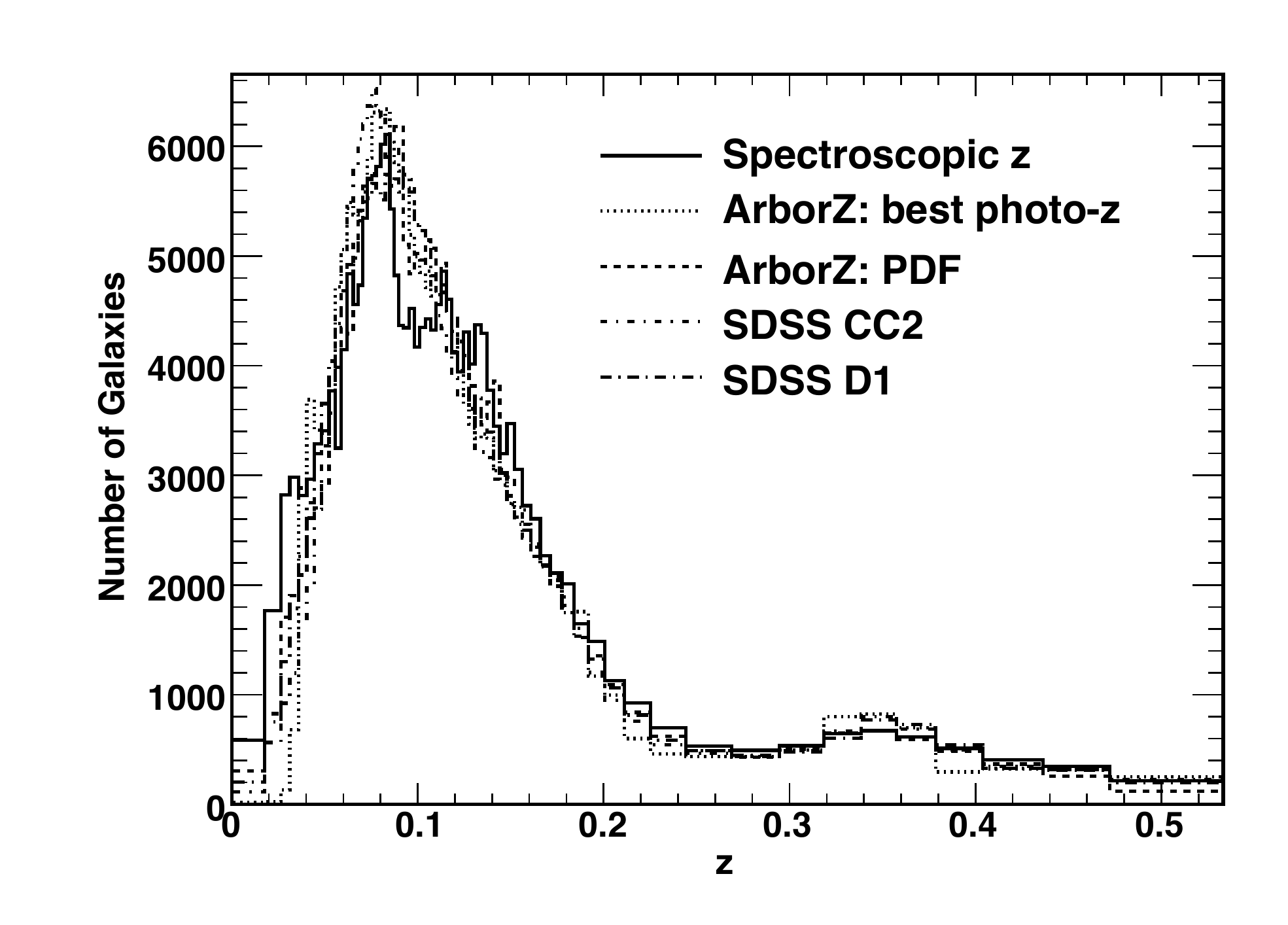} 
\includegraphics[height=\columnwidth,angle=90]{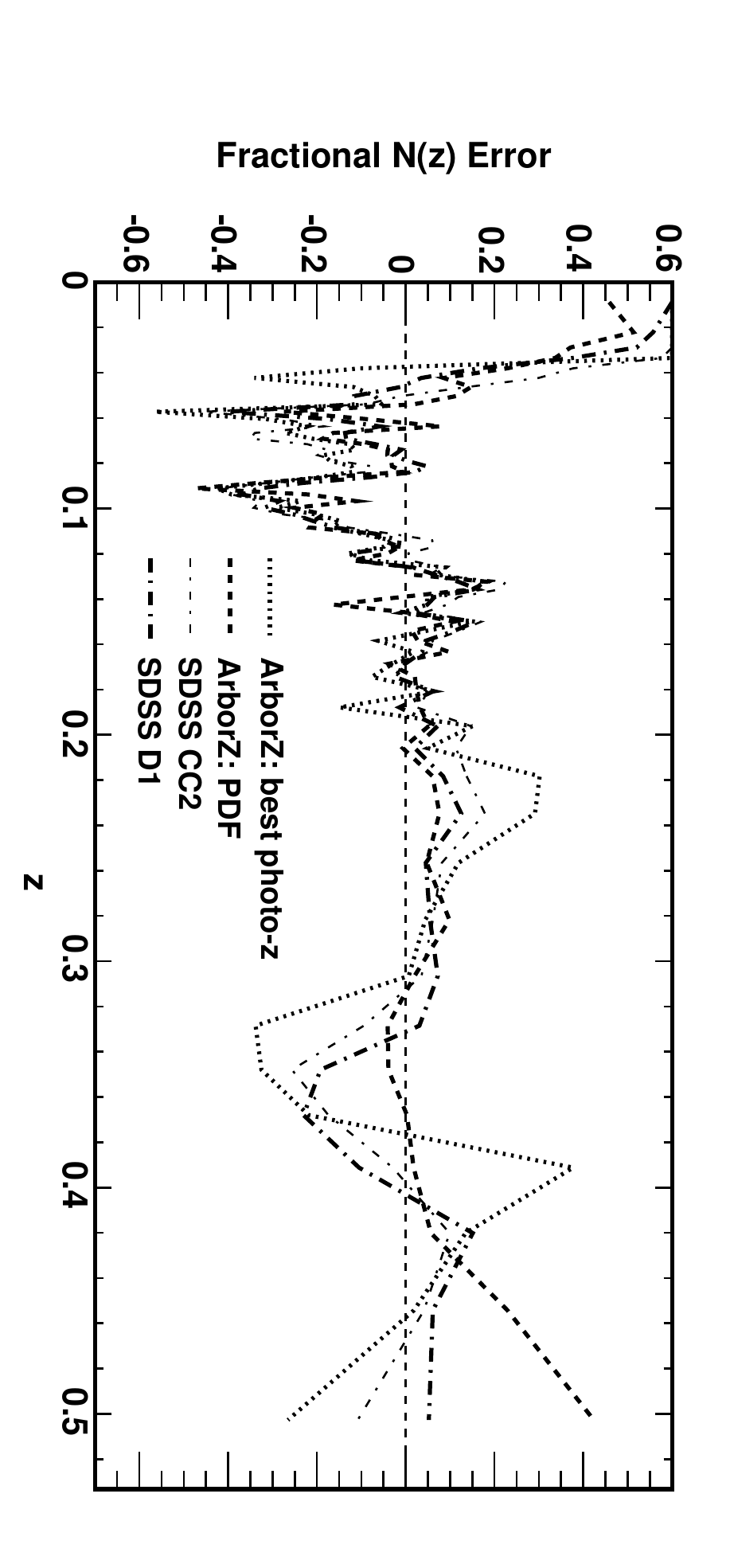}
 }
\caption{\label{fig:dNdz-SDSS} Top: Reconstructed redshift distribution in the SDSS spectroscopic evaluation  sample for four algorithms: ArborZ using the single best-estimate photo-$z$, the summed ArborZ PDFs, and the two SDSS production algorithms D1 and CC2. Bottom: Fractional error distribution ($N_{spec}-N_{photo})/N_{spec}$. }
\end{center}
\end{figure}

When normalized to unit area and corrected for the variable bin widths, the probability distributions illustrated in Figure~\ref{fig:BDToutputs} become PDFs. Figure~\ref{fig:dNdz-SDSS} shows the result of summing these PDFs for galaxies in the SDSS spectroscopic evaluation sample, together with the results from using the ArborZ  peak photo-$z$ estimate, and the two SDSS production photo-$z$ algorithms. As a quantitative comparison, we compute the goodness-of-reconstruction parameter
$$
	\chi^2 = \sum_i\left(\frac{N_{spec,i}-N_{photo,i}}{N_{spec,i}}\right)^2,
$$
where $i$ labels the redshift bins.
We find that $\chi^2 = 5.88, 4.04, 2.35,$~and 1.99 for the ArborZ (peak) method, CC2, D1, and ArborZ PDF method respectively. Thus, the summed PDFs provide the most faithfully reconstructed $N(z)$ of the algorithms considered.

\subsection{Performance in SDSS Mock Catalog}

\begin{figure}[htb]
\begin{center}
\ifthenelse{\boolean{ColorPlots}}
{\includegraphics[height=\columnwidth, angle=90]{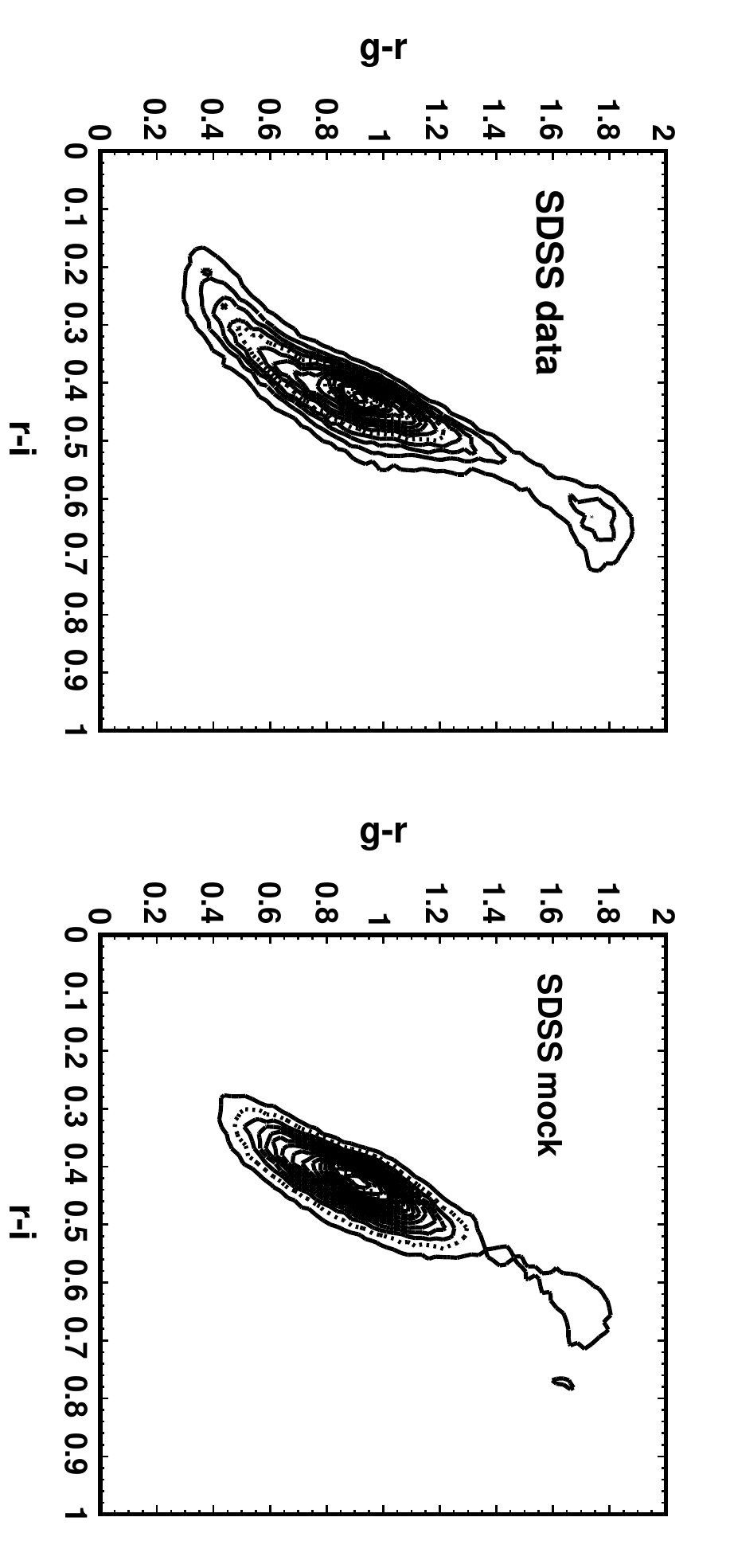} }
{\includegraphics[height=\columnwidth, angle=90]{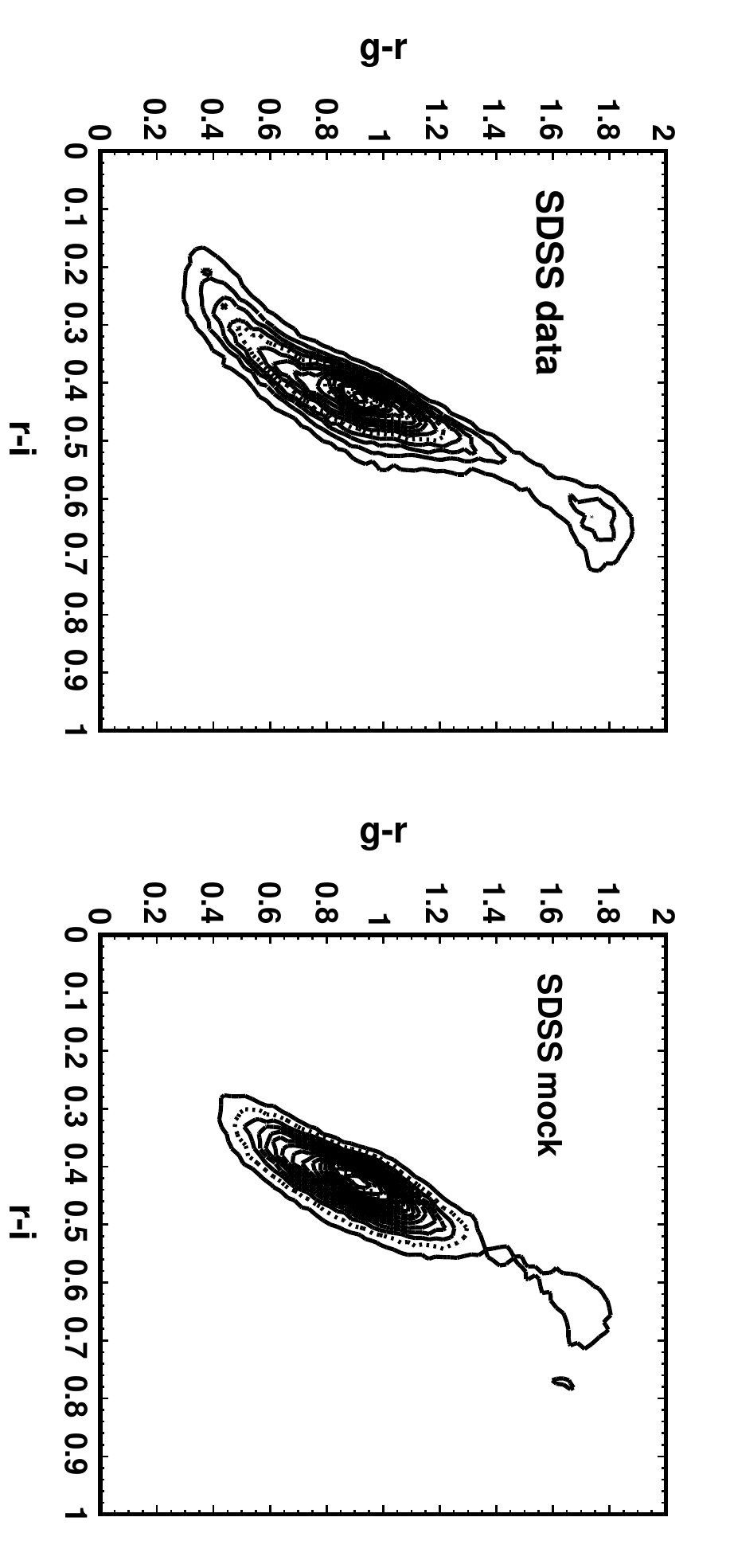} }
\caption{\label{fig:mockcolor} Color-color diagrams showing $g-r$ vs. $r-i$ for the SDSS spectroscopic sample (left) and the SDSS spectroscopic mock catalog (right). }
\end{center}
\end{figure}

Before characterizing the performance of the ArborZ algorithm in future surveys, we wish to establish the validity of our mock catalogs, described in Appendix~A, by constructing a mock catalog similar to the SDSS spectroscopic sample.
Using a larger mock with colors drawn from the full SDSS data sample, we simulate the SDSS spectroscopic selection to create a mock catalog containing both a low-redshift flux-limited component and a higher-redshift volume-limited population of LRGs. Color-color comparisons of the real SDSS spectroscopic sample and the mock catalog, such as the one shown in Figure~\ref{fig:mockcolor}, show good qualitative agreement. Differences in the outlier  populations are likely due to simplified treatment of SDSS photometric errors in  the mock catalog.

\begin{figure}[htb]
\begin{center}
\includegraphics[height=0.6\columnwidth, angle=90]{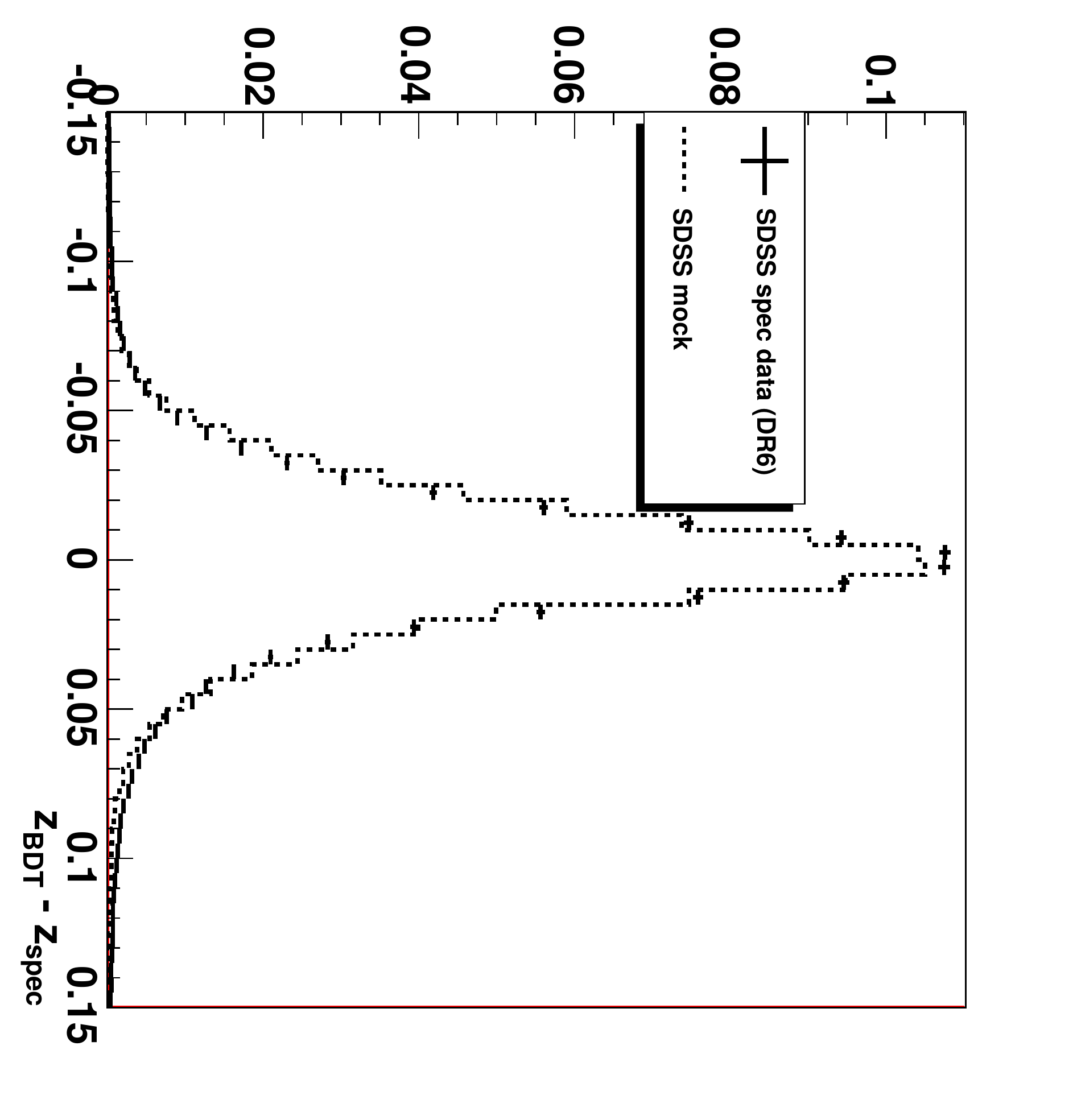}
\caption{\label{fig:mockerr} Comparison of the ArborZ photo-$z$ residual distribution in the SDSS mock spectroscopic catalog (histogram) and the real spectroscopic data (points).  }
\end{center}
\end{figure}

We train the ArborZ algorithm on the mock catalog using a training set identical in size to that employed in the real data. Figure~\ref{fig:mockerr} shows the photo-$z$ residual distribution for the SDSS spectroscopic mock, compared with the same distribution in real data, where the training in the data was also performed on the observed magnitudes only. The good agreement between these two distributions gives us confidence in extending these photo-$z$ error comparisons to mock catalogs for deeper surveys.

\subsection{Performance in DES Mock Catalog }
\label{sec:performance-DES}
\begin{figure}[htbp]
\begin{center}
\ifthenelse{\boolean{ColorPlots}}
{\includegraphics[width=0.7\columnwidth, angle=90]{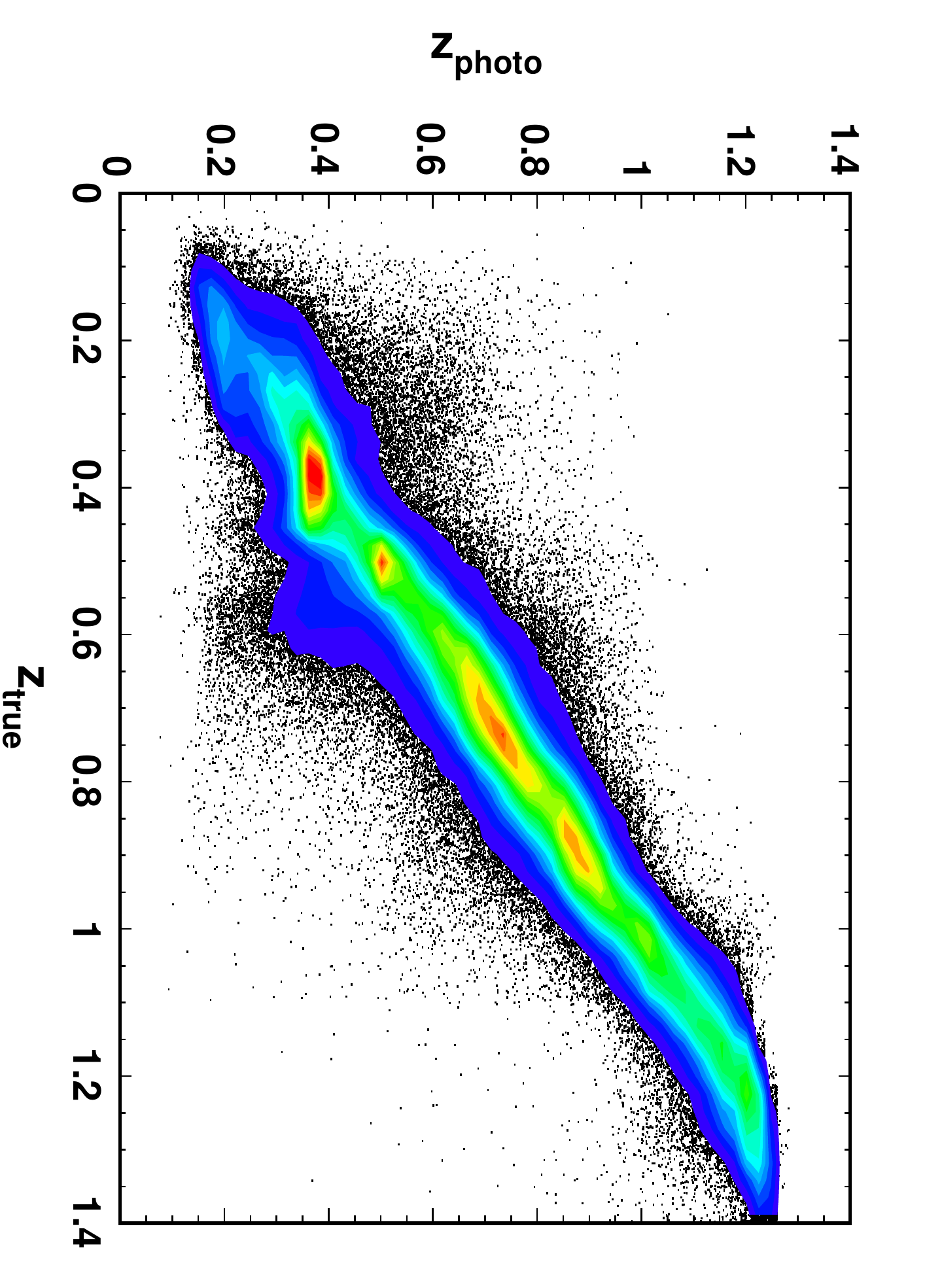}}
{\includegraphics[width=0.7\columnwidth, angle=90]{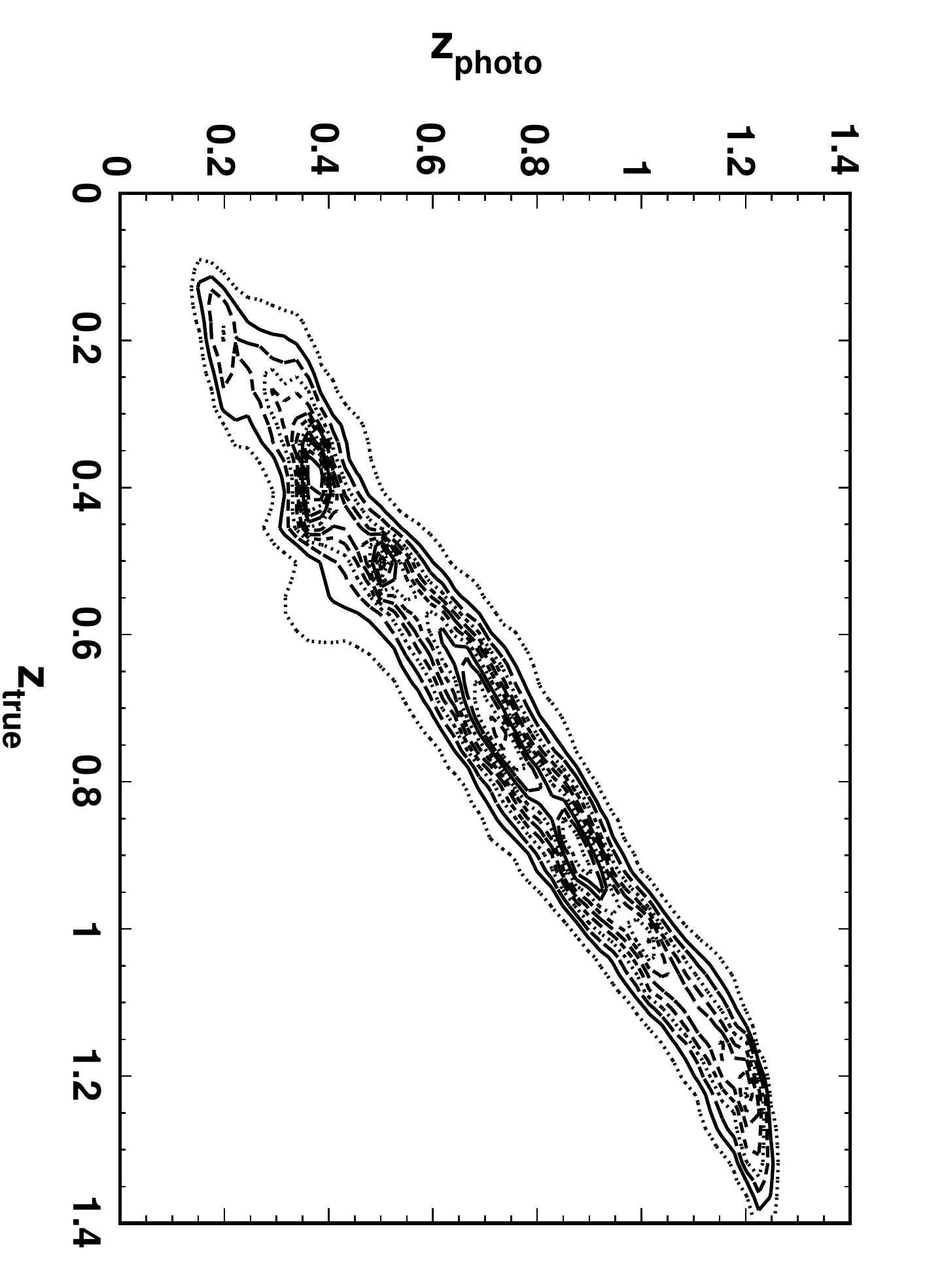}}
\caption{\label{fig:mockscatter} $z_{photo}$ vs. $z$ for galaxies in the DES mock catalog. }
\end{center}
\end{figure}

\begin{figure}[htbp]
\begin{center}
\includegraphics[height=\columnwidth,angle=90]{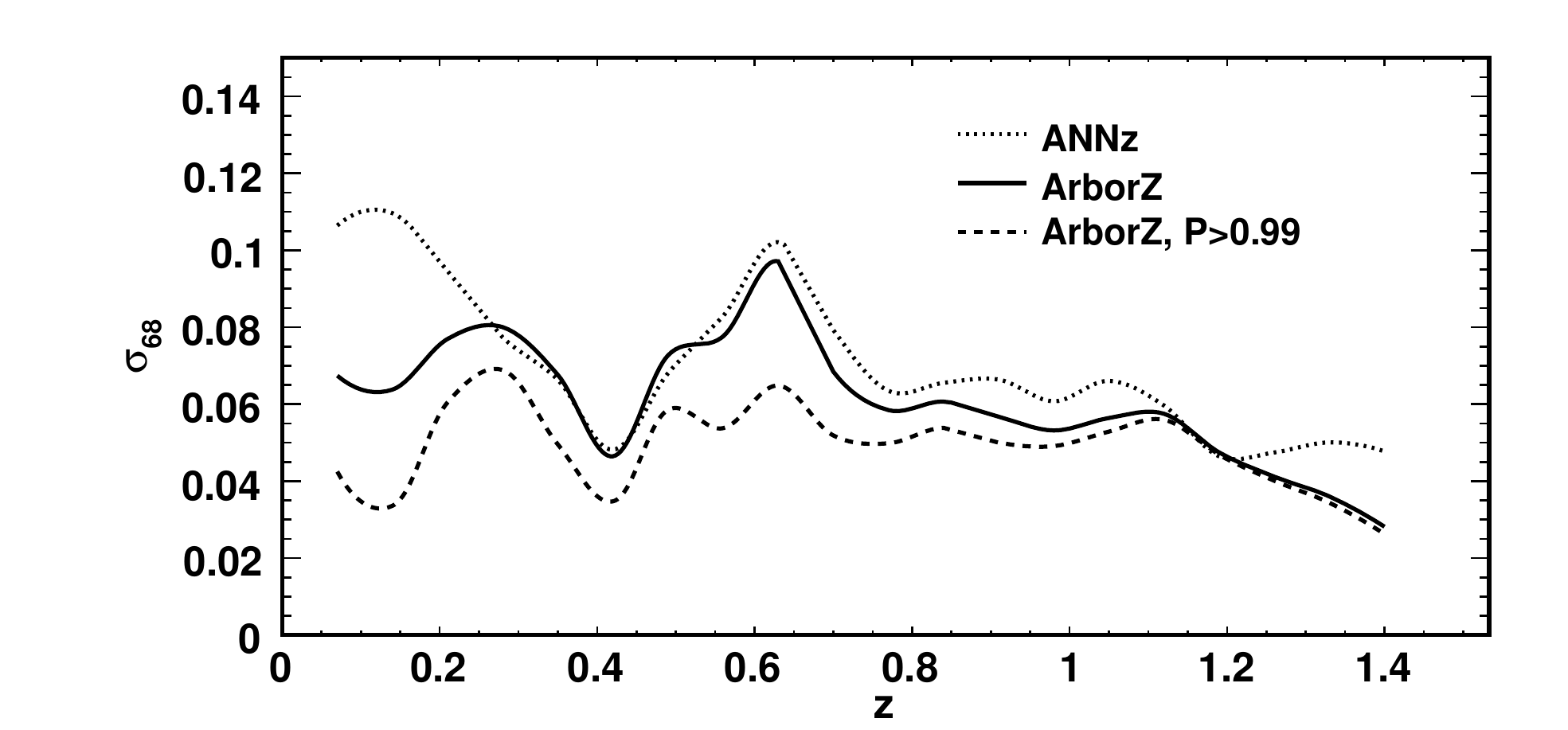}
\caption{\label{fig:sigzmock} Photo-$z$ error vs. $z$ for the DES mock catalog for the ArborZ algorithm (solid) and the neural net algorithm ANNz (dots). Also shown is the distribution for the 74\% of the galaxies in the catalog that pass a $P_{peak}>0.99$ cut.}
\end{center}
\end{figure}

\begin{figure}[htbp]
\begin{center}
\ifthenelse{\boolean{ColorPlots}}
{
\includegraphics[height=\columnwidth, angle=90]{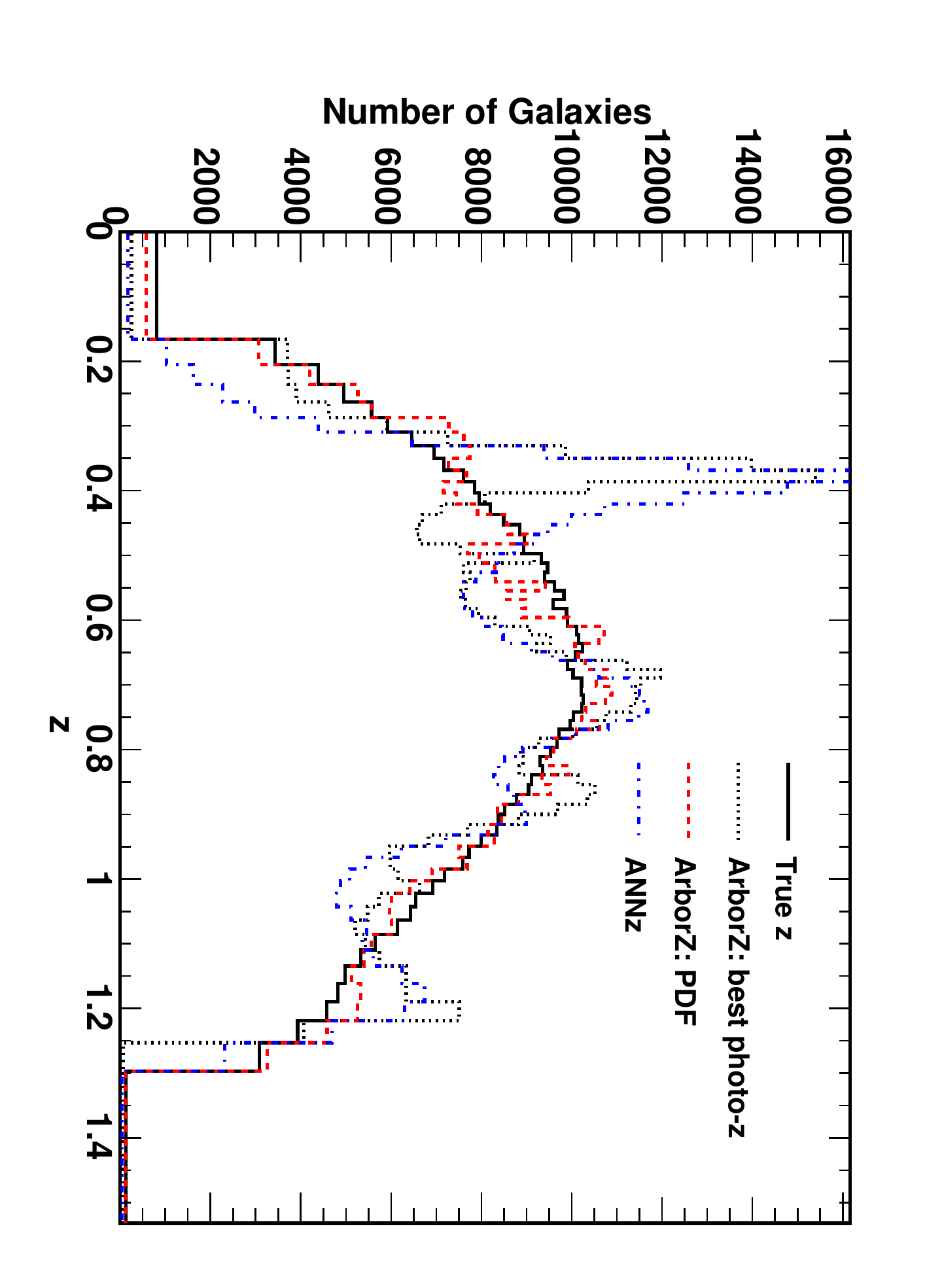}
\includegraphics[height=\columnwidth, angle=90]{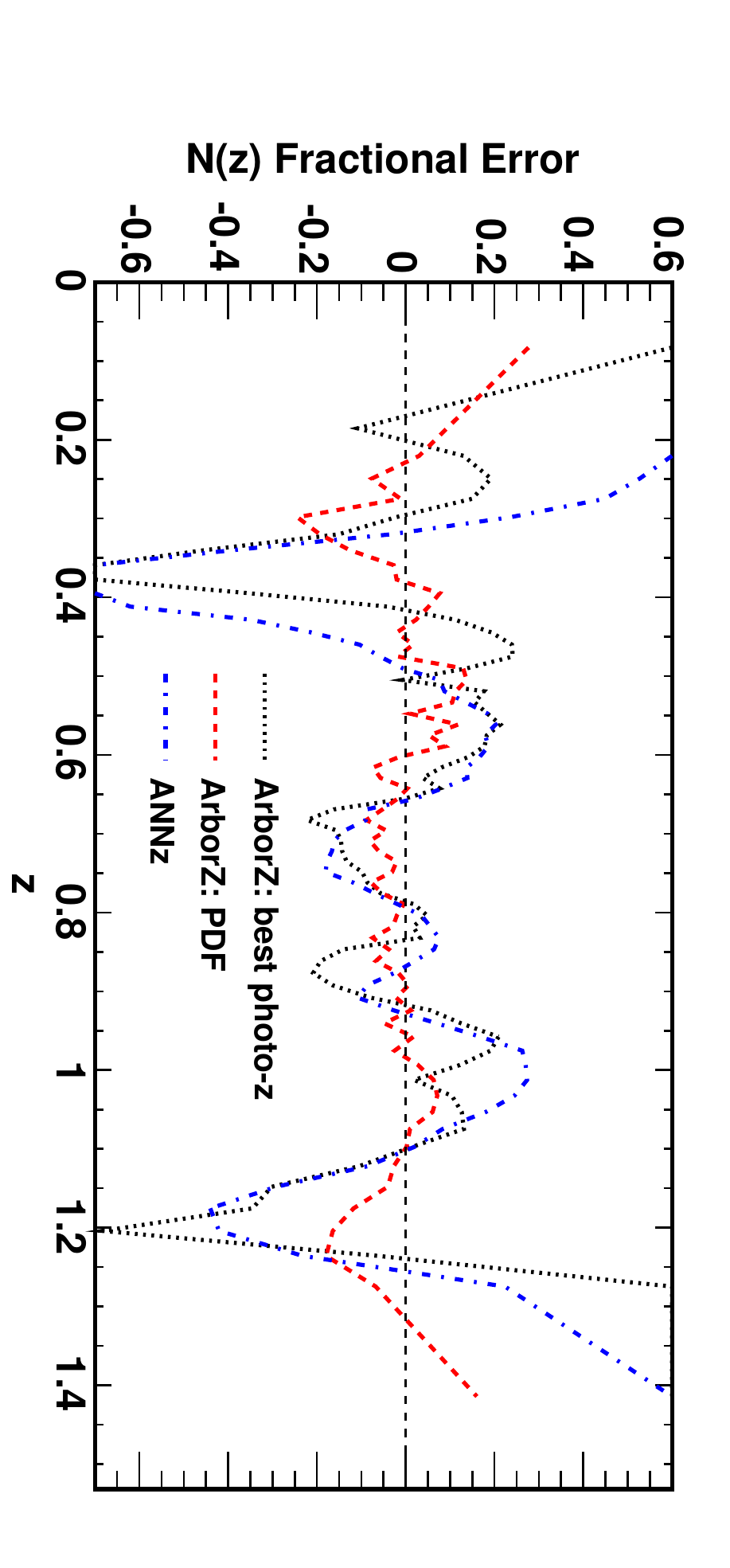}
}
{
\includegraphics[height=\columnwidth, angle=90]{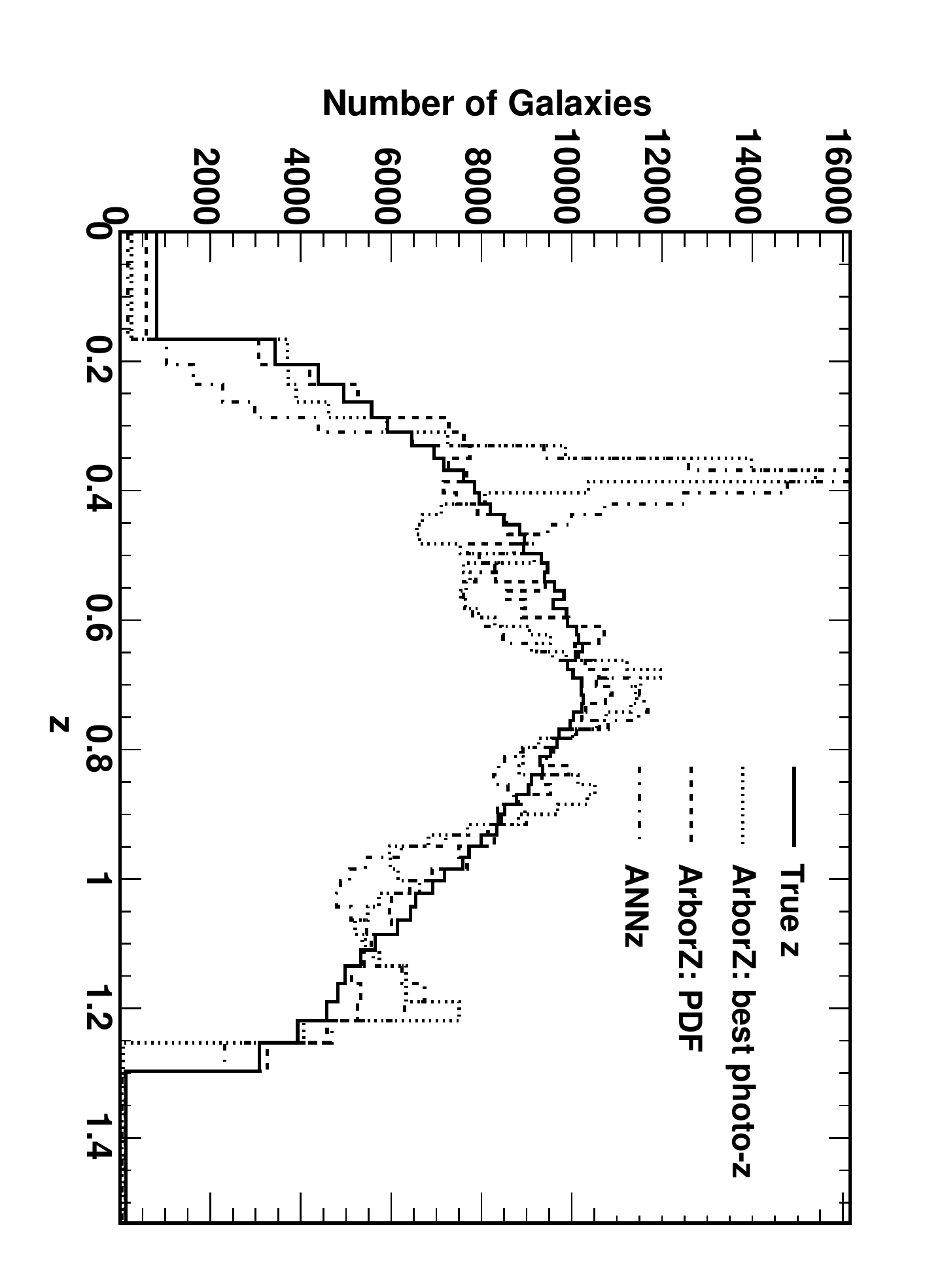}
\includegraphics[height=\columnwidth, angle=90]{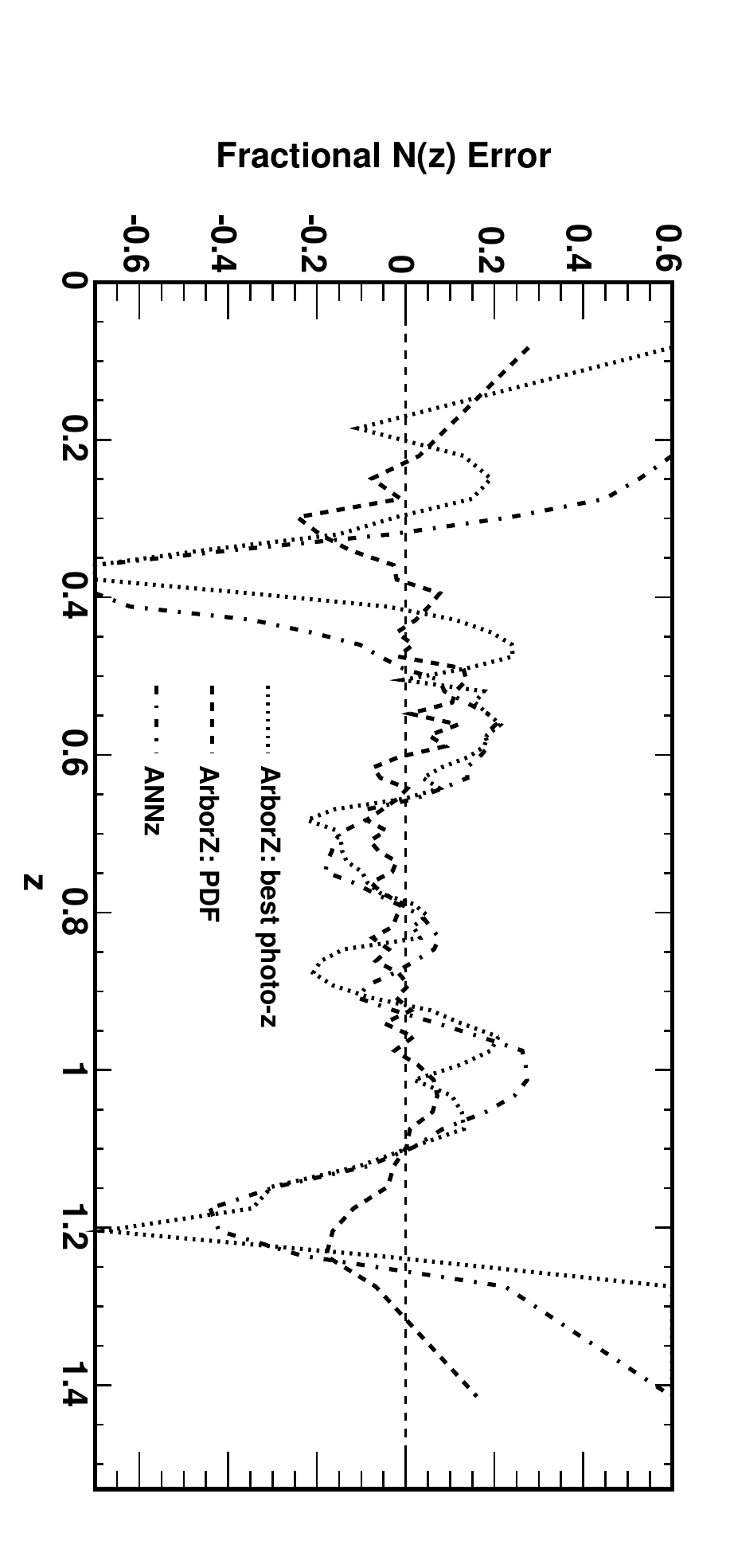}
}
\caption{\label{fig:zdist-DESmock} Top: Reconstructed redshift distributions in the DES mock catalog, compared with the true distribution, for the ArborZ (peak photoz), ANNz, and ArborZ (summed PDF) algorithms. Bottom: Fractional  error distribution ($N_{true}-N_{photo})/N_{true}$.  }
\end{center}
\end{figure}

Similarly, we have applied the ArborZ algorithm to the mock catalog of the Dark Energy Survey. We train and evaluate the algorithm using samples of 500k and 200k galaxies respectively, randomly selected from the full 20~million galaxy sample. We train on the observed $grizY$ magnitudes.  The resulting 
ArborZ photo-$z$ are shown in Figure ~\ref{fig:mockscatter}. For comparison, we have trained the neural net algorithm ANNz \citep{Collister2004, Firth2003} on the same training set. The neural net consists of a committee of five networks with five inputs (the observed magnitudes), two hidden layers with ten nodes each, and one output layer. The comparative performance of the two algorithms is illustrated in Figures~\ref{fig:sigzmock} and~\ref{fig:zdist-DESmock}. 

The redshift distributions $N(z)$ in Figure~\ref{fig:zdist-DESmock} display unphysical peaked structures when reconstructed by both the neural net and by ArborZ, when the peak photo-$z$ is used. This could indicate bias in the training, or possibly a problem with the color distribution in the mock catalog. However, the peaks largely disappear when the ArborZ PDFs are stacked to reconstruct the redshift distribution. The goodness-of-reconstruction parameter $\chi^2$ defined above is 7.10 for ANNz, 5.59 for the ArborZ (peak) method, and 0.45 using the ArborZ PDFs. The much better agreement obtained from using the PDFs highlights the limitations of using a single best-estimate photo-$z$ to characterize a galaxy, and shows the benefits of knowing the full PDF.

\section{Conclusions}
\label{sec:Conclusion}

We have presented a new technique to estimate photometric redshifts for galaxies. The technique, which we call ArborZ, uses Boosted Decision Tree classifiers trained on galaxies with spectroscopically-determined redshifts. In addition to providing a single best-estimate photo-$z$ with a reliably-calculated error, the method naturally produces a complete Probability Density Function for each galaxy's photo-$z$. The PDFs are shown to yield a more accurate reconstruction of the redshift distribution $N(z)$ than algorithms that rely on a single photo-$z$ for each galaxy. We also find that the peak probability for each galaxy provides a quantitative measure of the photo-$z$ quality, and can be used to define subsamples with better-measured photo-$z$s and fewer outliers. The performance of the algorithm on SDSS data with known spectroscopic redshifts in the range $0<z<0.5$ is comparable to or better than that of the production DR6 SDSS photo-$z$s, with slightly smaller errors and fewer catastrophic failures. We then studied the performance of the ArborZ algorithm on a simulated sample spanning a  much larger redshift range ($0<z<1.4$), intended to model the five-year sensitivity of the upcoming Dark Energy Survey. When trained on identical training sets, the ArborZ algorithm outperforms the artificial neural net algorithm ANNz, making it a promising candidate for determining photo-$z$s in deep photometric surveys.

\begin{comment}
We then investigated two examples of bias in photometric redshifts and explored methods for reducing it. First, we showed that standard photo-$z$s of highly inclined spiral galaxies are biased towards larger values due to reddening. However, by adding shape information as an additional training variable, this bias is significantly reduced. Turning to galaxy clusters, we showed that photo-$z$s of faint cluster members also tend to be biased toward higher values, because such objects are underrepresented in the spectroscopic training set. We remedied this by constructing a training set that includes cluster members drawn from MaxBCG clusters whose BCG has a spectroscopically-determined redshift, and assigning the member galaxies the same redshift as their corresponding BCG for training purposes. The resulting cluster photo-$z$s were found to be slightly better than those in the original MaxBCG catalog.
\end{comment}

As an empirical, learning-based algorithm, our approach does not provide a ready path to reliable photo-$z$s for objects significantly different from those in the training set. Future work will center upon exploring the biases inherent in different training sets, and on understanding the benefits and limitations of using simulated data to fill in gaps in these training sets.

\section{Acknowledgements}
This work is supported in part by the U.S. Department of Energy under grant DE-FG02-95ER40899, and by the National Science Foundations under grant AST 044327.
RHW and MB received support from the
U.S. Department of Energy under contract number DE-AC02-76SF00515
and from a Terman Fellowship at Stanford University.
% Standard Sloan acknowledgement
 Funding for the SDSS and SDSS-II has been provided by the Alfred P. Sloan Foundation, the Participating Institutions, the National Science Foundation, the U.S. Department of Energy, the National Aeronautics and Space Administration, the Japanese Monbukagakusho, the Max Planck Society, and the Higher Education Funding Council for England. The SDSS Web Site is http://www.sdss.org/.

    The SDSS is managed by the Astrophysical Research Consortium for the Participating Institutions. The Participating Institutions are the American Museum of Natural History, Astrophysical Institute Potsdam, University of Basel, University of Cambridge, Case Western Reserve University, University of Chicago, Drexel University, Fermilab, the Institute for Advanced Study, the Japan Participation Group, Johns Hopkins University, the Joint Institute for Nuclear Astrophysics, the Kavli Institute for Particle Astrophysics and Cosmology, the Korean Scientist Group, the Chinese Academy of Sciences (LAMOST), Los Alamos National Laboratory, the Max-Planck-Institute for Astronomy (MPIA), the Max-Planck-Institute for Astrophysics (MPA), New Mexico State University, Ohio State University, University of Pittsburgh, University of Portsmouth, Princeton University, the United States Naval Observatory, and the University of Washington.

\appendix
\section{Mock Catalogs and the ADDGALS Algorithm}
\label{sec:Addgals}

The Adding Density-Determined GAlaxies to Lightcone Simulations algorithm,
\citep[\addgals; ][]{Wechsler09}, is a method for producing
mock galaxy lightcone surveys that accurately reproduce the spatial
and color properties of a galaxy population.  It operates in
conjunction with a large volume, low mass-resolution $N$-body simulation,
adding galaxies with properties based on dark matter and galaxy
overdensities.  The first step of the algorithm includes one
observational input, the galaxy luminosity function in a given band.
Here we use the most recently published $r$-band luminosity function
measured from the SDSS \citep{Montero-Dorta08}, and assume passive
evolution of 1.3 mags per unit redshift \citep{Faber07}.  A list of
galaxies satisfying this luminosity function is generated and assigned
to dark matter particles in the simulation.  This is done using a
luminosity dependent function, $P(\delta | L_r/L_*)$, which specifies
the distribution of densities chosen as a function of galaxy
luminosity.  Here we use the dark matter density smoothed at the
Lagrangian scale corresponding to a mass near $M_*$, $M_{\rm smooth} =
1.8 \times 10^{13} \hMsun$.  The form of this PDF has been
determined by studying subhalos and semi-analytic galaxies in higher
resolution (but smaller volume) simulations.  We find that the radius
enclosing $M_{\rm smooth}$ is well represented by a lognormal plus a
Gaussian over a range of masses and luminosities.  In order to
determine the parameters of this PDF, specified by five parameters which
are each a function of $L_r/L_*$, we use a second observational input
as a constraint, the measured luminosity-dependent two-point
correlation function \citep{Zehavi05}. The $L_r/L_*$ term in the PDF
allows us to account for passive galaxy evolution but ignores other
evolutionary effects such as ongoing star formation.  This algorithm
is applied for galaxies brighter than $0.4L_*$.  The best-fit model
parameters are chosen using an MCMC analysis.  This results in a
distribution of galaxies with $r$-band magnitudes whose luminosity
function and clustering closely matches observations.  The use of the
smoothed background density instead of resolved halos allows the
method to populate large volume simulations with lower resolution than
would otherwise be possible, but sacrifices some fidelity in high
density regions.

One type of high density region which is not well-reproduced with this
approach is the centers of clusters, which host brightest cluster
galaxies (BCGs).  It has been shown that BCG luminosities are tightly
correlated with the masses of their host halos, and also that they are
brighter than would be indicated if they were selected from the
Schechter function of satellite galaxies in the same cluster
\citep{Hansen09}.  In order to account for these trends, we modify the
algorithm so that, before any other galaxies are inserted, a BCG
luminosity is calculated for each resolved halo of the simulation,
based on the observationally-constrained mean and scatter of the
luminosity--mass relationship for central galaxies \citep{Hansen09,
  Vale04, Zheng07}.  These objects are removed from our initial list
of galaxies and placed at the center of halos (in this catalog
version, this is done for halos more massive than $\sim 5 \times
10^{13} \hMsun$).

Once the simulation has been populated with galaxies with $r-$band
luminosities assigned, we assign colors to each galaxy in order to
mimic photometric surveys.  Our method assumes that galaxy colors are
set by both luminosity and environment.  We first compile a galaxy
training set from which we can measure the distribution of colors as a
function of luminosity and environment.  Here, we take the
magnitude-limited spectroscopic SDSS DR6 VAGC galaxy catalog and use
the density measurements of \citet*{Cooper08}.  The local galaxy
density is determined by calculating the projected distance to the
fifth nearest galaxy in a $\Delta z$ bin with velocity dispersion
width 1000 km~s$^{-1}$.  For each galaxy in our mock galaxy catalog, we
calculate the same density measurement, identify a SDSS galaxy with
similar density and $r-$band magnitude, and $k$-correct the colors of
this SDSS galaxy to the appropriate redshift for our mock galaxy.
When choosing densities, we consider {\sl relative} densities of
galaxies in each redshift bin, which mitigates differences between the
minimum magnitude used in calculating the densities between the
volume-limited mock and the magnitude-limited dataset.  We restrict
our SDSS sample to galaxies closer than $z < 0.2$ so that the bias of
observed galaxies remains relatively constant over this region.

When modeling deep surveys, at low redshifts we must add galaxies
dimmer than this algorithm can easily produce; the number density of
galaxies approaches (and, for the lowest redshifts, exceeds) the number
of dark matter particles in the simulation.  In this version of the
catalog, these galaxies (those dimmer than $0.4L_*(z)$) are placed
randomly in the volume, and should not be expected to have clustering
properties that match observed galaxies in detail.  In addition, the
limited depth of current surveys prevents us from compiling a training
set of colors for these galaxies.  We use the dimmest, bluest galaxies
in our training set, which extend down to absolute $r-$band magnitude
$\approx$ -15, and assume that these colors (which are likely somewhat too
red) are appropriate for the dimmest galaxies at low redshift.

This algorithm is very successful in reproducing photometric
properties for galaxies in the SDSS.  Here, we extend the algorithm to
substantially higher redshift and deeper depths to model DES.  There
are a number of issues in galaxy evolution that have not yet been
addressed in creating this catalog, and we highlight them here.  At high redshift we
simply extrapolate color information from low redshift galaxies
because of the lack of an appropriate training set.  In the present version
of the catalog, there is no stellar evolution modeling.  While galaxy
$r-$band magnitudes are passively evolved and spectra $k$-corrected,
we assume both that the typical rest-frame colors of galaxies are
unchanged and that the color-density-luminosity relation remains
unchanged.  Both of these assumptions are certainly incorrect in
detail.  Given these limitations, the detailed distribution of
photometric errors that come out of the algorithm at high redshift
should be treated with caution.  Future catalog versions will address
these issues.

 \bibliography{ArborZ}

\begin{thebibliography}{56}
\expandafter\ifx\csname natexlab\endcsname\relax\def\natexlab#1{#1}\fi

\bibitem[{{Abbott} {et~al.}(2005)}]{DESoverview}
{Abbott}, T. {et~al.} 2005, arXiv:astro-ph/0510346

\bibitem[{{Adelman-McCarthy} {et~al.}(2008)}]{Adelman-McCarthy08}
{Adelman-McCarthy}, J.~K. {et~al.} 2008, \apjs, 175, 297

\bibitem[{{Ball} {et~al.}(2008){Ball}, {Brunner}, {Myers}, {Strand}, {Alberts},
  \& {Tcheng}}]{Ball08}
{Ball}, N.~M., {Brunner}, R.~J., {Myers}, A.~D., {Strand}, N.~E., {Alberts},
  S.~L., \& {Tcheng}, D. 2008, \apj, 683, 12

\bibitem[{{Ball} {et~al.}(2007){Ball}, {Brunner}, {Myers}, {Strand}, {Alberts},
  {Tcheng}, \& {Llor{\`a}}}]{Ball07}
{Ball}, N.~M., {Brunner}, R.~J., {Myers}, A.~D., {Strand}, N.~E., {Alberts},
  S.~L., {Tcheng}, D., \& {Llor{\`a}}, X. 2007, \apj, 663, 774

\bibitem[{{Banerji} {et~al.}(2008){Banerji}, {Abdalla}, {Lahav}, \&
  {Lin}}]{Banerji08}
{Banerji}, M., {Abdalla}, F.~B., {Lahav}, O., \& {Lin}, H. 2008, \mnras, 386,
  1219

\bibitem[{{Baum}(1962)}]{baum62}
{Baum}, W.~A. 1962, in IAU Symposium, Vol.~15, Problems of Extra-Galactic
  Research, ed. G.~C. {McVittie}, 390

\bibitem[{{Ben{\'{\i}}tez}(2000)}]{Benitez00}
{Ben{\'{\i}}tez}, N. 2000, \apj, 536, 571

\bibitem[{{Blanton} {et~al.}(2003){Blanton}, {Lin}, {Lupton}, {Maley}, {Young},
  {Zehavi}, \& {Loveday}}]{Blanton2003}
{Blanton}, M.~R., {Lin}, H., {Lupton}, R.~H., {Maley}, F.~M., {Young}, N.,
  {Zehavi}, I., \& {Loveday}, J. 2003, \aj, 125, 2276

\bibitem[{{Breiman} {et~al.}(1984){Breiman}, {Friedman}, {Olshen}, \&
  {Stone}}]{Breiman1984}
{Breiman}, L., {Friedman}, J., {Olshen}, R., \& {Stone}, C. 1984,
  {Classification and Regression Trees} (Wadsworth International Group)

\bibitem[{{Bunn} \& {Hogg}(2009)}]{BunnHogg09}
{Bunn}, E.~F. \& {Hogg}, D.~W. 2009, American Journal of Physics, 77, 688

\bibitem[{{Cannon} {et~al.}(2006)}]{Cannon06}
{Cannon}, R. {et~al.} 2006, \mnras, 372, 425

\bibitem[{{Carliles} {et~al.}(2008){Carliles}, {Budav{\'a}ri}, {Heinis},
  {Priebe}, \& {Szalay}}]{Carliles08}
{Carliles}, S., {Budav{\'a}ri}, T., {Heinis}, S., {Priebe}, C., \& {Szalay}, A.
  2008, in Astronomical Society of the Pacific Conference Series, Vol. 394,
  Astronomical Data Analysis Software and Systems XVII, ed. R.~W. {Argyle},
  P.~S. {Bunclark}, \& J.~R. {Lewis}, 521--+

\bibitem[{{Coe} {et~al.}(2006){Coe}, {Ben{\'{\i}}tez}, {S{\'a}nchez}, {Jee},
  {Bouwens}, \& {Ford}}]{coe06}
{Coe}, D., {Ben{\'{\i}}tez}, N., {S{\'a}nchez}, S.~F., {Jee}, M., {Bouwens},
  R., \& {Ford}, H. 2006, \aj, 132, 926

\bibitem[{{Collister} \& {Lahav}(2004)}]{Collister2004}
{Collister}, A.~A. \& {Lahav}, O. 2004, \pasp, 116, 345

\bibitem[{{Connolly} {et~al.}(1995){Connolly}, {Csabai}, {Szalay}, {Koo},
  {Kron}, \& {Munn}}]{Connolly95}
{Connolly}, A.~J., {Csabai}, I., {Szalay}, A.~S., {Koo}, D.~C., {Kron}, R.~G.,
  \& {Munn}, J.~A. 1995, \aj, 110, 2655

\bibitem[{{Cooper} {et~al.}(2008){Cooper}, {Tremonti}, {Newman}, \&
  {Zabludoff}}]{Cooper08}
{Cooper}, M.~C., {Tremonti}, C.~A., {Newman}, J.~A., \& {Zabludoff}, A.~I.
  2008, \mnras, 390, 245

\bibitem[{{Coupon} {et~al.}(2009)}]{Coupon09}
{Coupon}, J. {et~al.} 2009, \aap, 500, 981

\bibitem[{{Cunha} {et~al.}(2009){Cunha}, {Lima}, {Oyaizu}, {Frieman}, \&
  {Lin}}]{Cunha09}
{Cunha}, C.~E., {Lima}, M., {Oyaizu}, H., {Frieman}, J., \& {Lin}, H. 2009,
  \mnras, 396, 2379

\bibitem[{{Davis} {et~al.}(2003)}]{Davis03}
{Davis}, M. {et~al.} 2003, in Society of Photo-Optical Instrumentation
  Engineers (SPIE) Conference Series, Vol. 4834, Society of Photo-Optical
  Instrumentation Engineers (SPIE) Conference Series, ed. P.~{Guhathakurta},
  161--172

\bibitem[{{Davis} {et~al.}(2007)}]{Davis07}
{Davis}, M. {et~al.} 2007, \apjl, 660, L1

\bibitem[{Drucker {et~al.}(1999)Drucker, Wu, \& Vapnik}]{Drucker1999}
Drucker, H., Wu, D., \& Vapnik, V. 1999, IEEE Transactions on Neural Networks,
  10, 1048

\bibitem[{{Eisenstein} {et~al.}(2001)}]{Eisenstein01}
{Eisenstein}, D.~J. {et~al.} 2001, \aj, 122, 2267

\bibitem[{{Faber} {et~al.}(2007)}]{Faber07}
{Faber}, S.~M. {et~al.} 2007, \apj, 665, 265

\bibitem[{{Feldmann} {et~al.}(2006)}]{Feldmann06}
{Feldmann}, R. {et~al.} 2006, \mnras, 372, 565

\bibitem[{{Fern{\'a}ndez-Soto} {et~al.}(1999){Fern{\'a}ndez-Soto}, {Lanzetta},
  \& {Yahil}}]{Fernandez99}
{Fern{\'a}ndez-Soto}, A., {Lanzetta}, K.~M., \& {Yahil}, A. 1999, \apj, 513, 34

\bibitem[{{Firth} {et~al.}(2003){Firth}, {Lahav}, \& {Somerville}}]{Firth2003}
{Firth}, A.~E., {Lahav}, O., \& {Somerville}, R.~S. 2003, \mnras, 339, 1195

\bibitem[{{Freeman} {et~al.}(2009){Freeman}, {Newman}, {Lee}, {Richards}, \&
  {Schafer}}]{Freeman09}
{Freeman}, P.~E., {Newman}, J.~A., {Lee}, A.~B., {Richards}, J.~W., \&
  {Schafer}, C.~M. 2009, \mnras, 1053

\bibitem[{{Freund} \& {Schapire}(1997)}]{Freund1997}
{Freund}, Y. \& {Schapire}, R.~E. 1997, JCSS, 55, 119

\bibitem[{{Giavalisco} {et~al.}(2004)}]{giav04}
{Giavalisco}, M. {et~al.} 2004, \apjl, 600, L93

\bibitem[{{Hansen} {et~al.}(2009){Hansen}, {Sheldon}, {Wechsler}, \&
  {Koester}}]{Hansen09}
{Hansen}, S.~M., {Sheldon}, E.~S., {Wechsler}, R.~H., \& {Koester}, B.~P. 2009,
  \apj, 699, 1333

\bibitem[{{Hastie} {et~al.}(2001){Hastie}, {Tibshirani}, \&
  {Friedman}}]{Hastie2001}
{Hastie}, R., {Tibshirani}, R., \& {Friedman}, J. 2001, {The Elements of
  Statistical Learning: Data Mining, Inference, and Prediction} (Springer
  Series in Statistics)

\bibitem[{{Hoecker} {et~al.}(2007)}]{Hocker2007}
{Hoecker}, A. {et~al.} 2007, arXiv:physics/0703039

\bibitem[{Howe {et~al.}(2005)Howe, Rath, \& Manmatha}]{Howe2005}
Howe, N.~R., Rath, T.~M., \& Manmatha, R. 2005, in SIGIR '05: Proceedings of
  the 28th annual international ACM SIGIR conference on Research and
  development in information retrieval (New York, NY, USA: ACM), 377--383

\bibitem[{{Ilbert} {et~al.}(2006)}]{Ilbert06}
{Ilbert}, O. {et~al.} 2006, \aap, 457, 841

\bibitem[{{Ilbert} {et~al.}(2009)}]{Ilbert09}
---. 2009, \apj, 690, 1236

\bibitem[{{Ivezic} {et~al.}(2008)}]{ive08}
{Ivezic}, Z. {et~al.} 2008, ArXiv e-prints

\bibitem[{{Koo}(1985)}]{koo85}
{Koo}, D.~C. 1985, \aj, 90, 418

\bibitem[{{Koo}(1981)}]{koo81}
{Koo}, D.~C.-Y. 1981, PhD thesis, AA(University of California, Berkeley.)

\bibitem[{{Lima} {et~al.}(2008){Lima}, {Cunha}, {Oyaizu}, {Frieman}, {Lin}, \&
  {Sheldon}}]{Lima08}
{Lima}, M., {Cunha}, C.~E., {Oyaizu}, H., {Frieman}, J., {Lin}, H., \&
  {Sheldon}, E.~S. 2008, \mnras, 390, 118

\bibitem[{{Lin} {et~al.}(2004){Lin}, {Cunha}, {Lima}, {Oyaizu}, {Frieman},
  {Collister}, {Lahav}, \& {Dark Energy Survey}}]{Lin04}
{Lin}, H., {Cunha}, C., {Lima}, M., {Oyaizu}, H., {Frieman}, J., {Collister},
  A., {Lahav}, O., \& {Dark Energy Survey}. 2004, in Bulletin of the American
  Astronomical Society, Vol.~36, Bulletin of the American Astronomical Society,
  1462

\bibitem[{{Loh} \& {Spillar}(1986{\natexlab{a}})}]{lohsp86-2}
{Loh}, E.~D. \& {Spillar}, E.~J. 1986{\natexlab{a}}, \apjl, 307, L1

\bibitem[{{Loh} \& {Spillar}(1986{\natexlab{b}})}]{lohsp86}
---. 1986{\natexlab{b}}, \apj, 303, 154

\bibitem[{{Mandelbaum} {et~al.}(2008)}]{Mandelbaum08}
{Mandelbaum}, R. {et~al.} 2008, \mnras, 386, 781

\bibitem[{{Mitchell}(1997)}]{mit97}
{Mitchell}, T.~M. 1997, {Machine Learning} (McGraw-Hill)

\bibitem[{{Montero-Dorta} \& {Prada}(2008)}]{Montero-Dorta08}
{Montero-Dorta}, A.~D. \& {Prada}, F. 2008, ArXiv e-prints

\bibitem[{{Newman}(2008)}]{new08}
{Newman}, J.~A. 2008, \apj, 684, 88

\bibitem[{{Oyaizu} {et~al.}(2008){Oyaizu}, {Lima}, {Cunha}, {Lin}, {Frieman},
  \& {Sheldon}}]{Oyaizu2008}
{Oyaizu}, H., {Lima}, M., {Cunha}, C.~E., {Lin}, H., {Frieman}, J., \&
  {Sheldon}, E.~S. 2008, \apj, 674, 768

\bibitem[{{Roe} {et~al.}(2005){Roe}, {Yang}, {Zhu}, {Liu}, {Stancu}, \&
  {McGregor}}]{Roe2005}
{Roe}, B.~P., {Yang}, H.-J., {Zhu}, J., {Liu}, Y., {Stancu}, I., \& {McGregor},
  G. 2005, Nuclear Instruments and Methods in Physics Research A, 543, 577

\bibitem[{{Scoville} {et~al.}(2007)}]{sco07}
{Scoville}, N. {et~al.} 2007, \apjs, 172, 1

\bibitem[{{Strauss} {et~al.}(2002)}]{Strauss2002}
{Strauss}, M.~A. {et~al.} 2002, \aj, 124, 1810

\bibitem[{{Vale} \& {Ostriker}(2004)}]{Vale04}
{Vale}, A. \& {Ostriker}, J.~P. 2004, \mnras, 353, 189

\bibitem[{{Vanzella} {et~al.}(2004)}]{Vanzella04}
{Vanzella}, E. {et~al.} 2004, \aap, 423, 761

\bibitem[{{Wechsler} {et~al.}(2009)}]{Wechsler09}
{Wechsler}, R.~H. {et~al.} 2009, in preparation

\bibitem[{{York} {et~al.}(2000)}]{SDSSoverview}
{York}, D.~G. {et~al.} 2000, \aj, 120, 1579

\bibitem[{{Zehavi} {et~al.}(2005)}]{Zehavi05}
{Zehavi}, I. {et~al.} 2005, \apj, 630, 1

\bibitem[{{Zheng} {et~al.}(2007){Zheng}, {Coil}, \& {Zehavi}}]{Zheng07}
{Zheng}, Z., {Coil}, A.~L., \& {Zehavi}, I. 2007, \apj, 667, 760

\end{thebibliography}

\label{lastpage}
\end{document}